\documentclass[]{aastex63}
\usepackage[english]{babel} 
\usepackage{amssymb}
\usepackage{amsmath}
\usepackage{txfonts}
\usepackage{mathdots}
\usepackage[classicReIm]{kpfonts}
\usepackage{graphicx}
\usepackage{subfigure}
\usepackage{multirow}
\usepackage{aas_macros}
\usepackage{textcomp}
\usepackage[bottom]{footmisc}

\received{\today} \revised{\today} \accepted{\today}
\submitjournal{AJ}

\shorttitle{IXPE Instrument} \shortauthors{Soffitta et al.} 

\begin{document}

\title{The Instrument of the Imaging X-ray Polarimetry Explorer \footnote{Released on 2021}}

\correspondingauthor{Paolo Soffitta}
\email{paolo.soffitta@inaf.it}

\author{Paolo Soffitta}
\affiliation{Istituto di Astrofisica e Planetologia Spaziali, Via
del Fosso del Cavaliere 100, 00133 Rome, Italy}
\author{Luca Baldini}
\affiliation{Universita' di Pisa, Lungarno Pacinotti 43, 56126
Pisa} \affiliation{Istituto Nazionale di Fisica Nucleare, Largo B.
Pontecorvo 3, 56127 Pisa, Italy}
\author{Ronaldo Bellazzini}
\affiliation{Istituto Nazionale di Fisica Nucleare, Largo B.
Pontecorvo 3, 56127 Pisa, Italy}
\author{Enrico Costa}
\affiliation{Istituto di Astrofisica e Planetologia Spaziali, Via
del Fosso del Cavaliere 100, 00133 Rome, Italy}
\author{Luca Latronico}
\affiliation{Istituto Nazionale di Fisica Nucleare, Via P. Giuria
1, 10125  Turin, Italy}
\author{Fabio Muleri}
\affiliation{Istituto di Astrofisica e Planetologia Spaziali, Via
del Fosso del Cavaliere 100, 00133 Rome, Italy}
\author{Ettore Del Monte}
\author{Sergio Fabiani}
\affiliation{Istituto di Astrofisica e Planetologia Spaziali, Via
del Fosso del Cavaliere 100, 00133 Rome, Italy}
\author{Massimo Minuti}
\author{Michele Pinchera}
\author{Carmelo Sgro'}
\author{Gloria Spandre}
\affiliation{Istituto Nazionale di Fisica Nucleare, Largo B.
Pontecorvo 3, 56127 Pisa, Italy}
\author{Alessio Trois}
\affiliation{Osservatorio Astronomico di Cagliari, Via della
Scienza 5, 09047 Selargius (Ca) Italy}
\author{Fabrizio Amici}
\affiliation{Istituto di Astrofisica e Planetologia Spaziali, Via
del Fosso del Cavaliere 100, 00133 Rome, Italy}
\author{Hans Andersson}
\affiliation{Oxford Instruments Technologies  Oy, Technopolis
Innopoli 1, Tekniikantie 12, FI-02150 Espoo, Finland}
\author{Primo Attina'}
\affiliation{Osservatorio Astrofisico di Torino Via Osservatorio,
20, 10025 Pino Torinese (Turin), Italy}
\author{Matteo Bachetti}
\affiliation{Osservatorio Astronomico di Cagliari, Via della
Scienza 5, 09047 Selargius (Ca) Italy}
\author{Mattia Barbanera}
\affiliation{Istituto Nazionale di Fisica Nucleare, Largo B.
Pontecorvo 3, 56127 Pisa, Italy}
\author{Fabio Borotto}
\affiliation{Istituto Nazionale di Fisica Nucleare, Via P. Giuria
1, 10125  Turin, Italy}
\author{Alessandro Brez}
\affiliation{Istituto Nazionale di Fisica Nucleare, Largo B.
Pontecorvo 3, 56127 Pisa, Italy}
\author {Daniele Brienza}
\affiliation{Istituto di Astrofisica e Planetologia Spaziali, Via
del Fosso del Cavaliere 100, 00133 Rome, Italy}
\author{Ciro Caporale}
\affiliation{Istituto Nazionale di Fisica Nucleare, Via P. Giuria
1, 10125  Turin, Italy}
\author{Claudia Cardelli}
\affiliation{Istituto Nazionale di Fisica Nucleare, Largo B.
Pontecorvo 3, 56127 Pisa, Italy}
\author{Rita Carpentiero}
\affiliation{Agenzia Spaziale Italiana Via del Politecnico snc,
00133 Rome, Italy}
\author{Simone Castellano}
\affiliation{Istituto Nazionale di Fisica Nucleare, Largo B.
Pontecorvo 3, 56127 Pisa, Italy}
\author{Marco Castronuovo}
\affiliation{Agenzia Spaziale Italiana Via del Politecnico snc,
00133 Rome, Italy}
\author{Luca Cavalli}
\affiliation{Orbitale Hochtechnologie Bremen, OHB Italia, Via
Gallarate 150, 20151 Milan Italy}
\author{Elisabetta Cavazzuti}
\affiliation{Agenzia Spaziale Italiana Via del Politecnico snc,
00133 Rome, Italy}
\author{Marco Ceccanti}
\affiliation{Istituto Nazionale di Fisica Nucleare, Largo B.
Pontecorvo 3, 56127 Pisa, Italy}
\author{Mauro Centrone}
\affiliation{Osservtorio Astronomico di Rome, Via Frascati 33,
00078 Monte Porzio Catone (Rome) Italy}
\author{Stefano Ciprini}
\affiliation{Istituto Nazionale di Fisica Nucleare (INFN) Sezione
di Rome Tor Vergata, Via della Ricerca Scientifica 1, 00133 Rome,
Italy} \affiliation{Agenzia Spaziale Italiana Space Science Data
Center (SSDC) Via del Politecnico snc, 00133 Rome, Italy}
\author{Saverio Citraro}
\affiliation{Istituto Nazionale di Fisica Nucleare, Largo B.
Pontecorvo 3, 56127 Pisa, Italy}
\author{Fabio D'Amico}
\affiliation{Agenzia Spaziale Italiana Via del Politecnico snc,
00133 Rome, Italy}
\author{Elisa D'Alba}
\affiliation{Orbitale Hochtechnologie Bremen, OHB Italia, Via
Gallarate 150, 20151 Milan Italy}
\author{Sergio Di Cosimo}
\affiliation{Istituto di Astrofisica e Planetologia Spaziali, Via
del Fosso del Cavaliere 100, 00133 Rome, Italy}
\author{Niccolo' Di Lalla}
\affiliation{W.  W.  Hansen  Experimental  Physics  Laboratory,
Kavli  Institute  for  Particle  Astro-physics and Cosmology,
Department of Physics and SLAC National Accelerator
Laboratory,Stanford University, Stanford, CA 94305, USA}
\author{Alessandro Di Marco}
\affiliation{Istituto di Astrofisica e Planetologia Spaziali, Via
del Fosso del Cavaliere 100, 00133 Rome, Italy}
\author{Giuseppe Di Persio}
\affiliation{Istituto di Astrofisica e Planetologia Spaziali, Via
del Fosso del Cavaliere 100, 00133 Rome, Italy}
\author{Immacolata Donnarumma}
\affiliation{Agenzia Spaziale Italiana Via del Politecnico snc,
00133 Rome, Italy}
\author{Yuri Evangelista}
\affiliation{Istituto di Astrofisica e Planetologia Spaziali, Via
del Fosso del Cavaliere 100, 00133 Rome, Italy}
\author{Riccardo Ferrazzoli}
\affiliation{Istituto di Astrofisica e Planetologia Spaziali, Via
del Fosso del Cavaliere 100, 00133 Rome, Italy}
\author{Asami Hayato}
\affiliation{RIKEN Nishina Center, 2-1 Hirosawa, Wako, Saitama
351-0198, Japan}
\author{Takao Kitaguchi}
\affiliation{RIKEN Nishina Center, 2-1 Hirosawa, Wako, Saitama
351-0198, Japan}
\author{Fabio La Monaca}
\affiliation{Istituto di Astrofisica e Planetologia Spaziali, Via
del Fosso del Cavaliere 100, 00133 Rome, Italy}
\author{Carlo Lefevre}
\affiliation{Istituto di Astrofisica e Planetologia Spaziali, Via
del Fosso del Cavaliere 100, 00133 Rome, Italy}
\author{Pasqualino Loffredo} \affiliation{Istituto di Astrofisica e
Planetologia Spaziali, Via del Fosso del Cavaliere 100, 00133
Rome, Italy}
\author{Paolo Lorenzi}
\affiliation{Orbitale Hochtechnologie Bremen, OHB Italia, Via
Gallarate 150, 20151 Milan Italy}
\author{Leonardo Lucchesi}
\affiliation{Istituto Nazionale di Fisica Nucleare, Largo B.
Pontecorvo 3, 56127 Pisa, Italy}
\author{Carlo Magazzu}
\affiliation{Istituto Nazionale di Fisica Nucleare, Largo B.
Pontecorvo 3, 56127 Pisa, Italy}
\author{Simone Maldera}
\affiliation{Istituto Nazionale di Fisica Nucleare, Via P. Giuria
1, 10125  Turin, Italy}
\author{Alberto Manfreda}
\affiliation{Istituto Nazionale di Fisica Nucleare, Largo B.
Pontecorvo 3, 56127 Pisa, Italy}
\author{Elio Mangraviti}
\affiliation{Orbitale Hochtechnologie Bremen, OHB Italia, Via
Gallarate 150, 20151 Milan Italy}
\author{Marco Marengo}
\affiliation{Istituto Nazionale di Fisica Nucleare, Via P. Giuria
1, 10125  Turin, Italy}
\author{Giorgio Matt}
\affiliation{Dipartimento di Matematica e Fisica, Universita'
degli Studi Rome Tre, Via della Vasca Navale 84, 00146 Rome,
Italy}
\author{Paolo Mereu}
\affiliation{Istituto Nazionale di Fisica Nucleare, Via P. Giuria
1, 10125  Turin, Italy}
\author{Alfredo Morbidini}
\affiliation{Istituto di Astrofisica e Planetologia Spaziali, Via
del Fosso del Cavaliere 100, 00133 Rome, Italy}
\author{Federico Mosti}
\affiliation{Istituto Nazionale di Fisica Nucleare, Via P. Giuria
1, 10125  Turin, Italy}
\author{Toshio Nakano}
\affiliation{RIKEN Nishina Center, 2-1 Hirosawa, Wako, Saitama
351-0198, Japan}
\author{Hikmat Nasimi}
\affiliation{Istituto di Astrofisica e Planetologia Spaziali, Via
del Fosso del Cavaliere 100, 00133 Rome, Italy}
\author{Barbara Negri}
\affiliation{Agenzia Spaziale Italiana Via del Politecnico snc,
00133 Rome, Italy}
\author{Seppo Nenonen}
\affiliation{Oxford Instruments Technologies  Oy, Technopolis
Innopoli 1, Tekniikantie 12, FI-02150 Espoo, Finland}
\author{Alessio Nuti}
\affiliation{Istituto Nazionale di Fisica Nucleare, Largo B.
Pontecorvo 3, 56127 Pisa, Italy}
\author{Leonardo Orsini}
\affiliation{Istituto Nazionale di Fisica Nucleare, Largo B.
Pontecorvo 3, 56127 Pisa, Italy}
\author{Matteo Perri}
\affiliation{Osservtorio Astronomico di Rome, Via Frascati 33,
00078 Monte Porzio Catone (Rome) Italy}
\author{Melissa Pesce-Rollins}
\affiliation{Istituto Nazionale di Fisica Nucleare, Largo B.
Pontecorvo 3, 56127 Pisa, Italy}
\author{Raffaele Piazzolla}
\affiliation{Istituto di Astrofisica e Planetologia Spaziali, Via
del Fosso del Cavaliere 100, 00133 Rome, Italy}
\author{Maura Pilia}
\affiliation{Osservatorio Astronomico di Cagliari, Via della
Scienza 5, 09047 Selargius (Ca) Italy}
\author{Alessandro Profeti}
\affiliation{Istituto Nazionale di Fisica Nucleare, Largo B.
Pontecorvo 3, 56127 Pisa, Italy}
\author{Simonetta Puccetti}
\affiliation{Agenzia Spaziale Italiana Via del Politecnico snc,
00133 Rome, Italy}
\author{John Rankin}
\affiliation{Istituto di Astrofisica e Planetologia Spaziali, Via
del Fosso del Cavaliere 100, 00133 Rome, Italy}
\author{Ajay Ratheesh}
\affiliation{Istituto di Astrofisica e Planetologia Spaziali, Via
del Fosso del Cavaliere 100, 00133 Rome, Italy}
\affiliation{Dipartimento Di Fisica Universita' degli Studi di
Rome "Tor Vergata", Via della Ricerca Scientifica 1, 00133 Rome}
\author{Alda Rubini}
\affiliation{Istituto di Astrofisica e Planetologia Spaziali, Via
del Fosso del Cavaliere 100, 00133 Rome, Italy}
\author{Francesco Santoli}
\affiliation{Istituto di Astrofisica e Planetologia Spaziali, Via
del Fosso del Cavaliere 100, 00133 Rome, Italy}
\author{Paolo Sarra}
\affiliation{Orbitale Hochtechnologie Bremen, OHB Italia, Via
Gallarate 150, 20151 Milan Italy}
\author{Emanuele Scalise}
\affiliation{Istituto di Astrofisica e Planetologia Spaziali, Via
del Fosso del Cavaliere 100, 00133 Rome, Italy}
\author{Andrea Sciortino}
\affiliation{Orbitale Hochtechnologie Bremen, OHB Italia, Via
Gallarate 150, 20151 Milan Italy}
\author{Toru Tamagawa}
\affiliation{RIKEN Nishina Center, 2-1 Hirosawa, Wako, Saitama
351-0198, Japan}
\author{Marcello Tardiola}
\affiliation{Istituto Nazionale di Fisica Nucleare, Via P. Giuria
1, 10125  Turin, Italy}
\author{Antonino Tobia}
\affiliation{Istituto di Astrofisica e Planetologia Spaziali, Via
del Fosso del Cavaliere 100, 00133 Rome, Italy}
\author{Marco Vimercati}
\affiliation{Orbitale Hochtechnologie Bremen, OHB Italia, Via
Gallarate 150, 20151 Milan Italy}
\author{Fei Xie}
\affiliation{Istituto di Astrofisica e Planetologia Spaziali, Via
del Fosso del Cavaliere 100, 00133 Rome, Italy}



\begin{abstract}
While X-ray  Spectroscopy, Timing and Imaging have improved very
much since 1962, when the first astronomical non-solar source was
discovered, especially with the launch of Newton/X-ray
Multi-Mirror Mission, Rossi/X-ray Timing Explorer and
Chandra/Advanced X-ray Astrophysics Facility, the progress of
X-ray polarimetry has been meager. This is in part due to the lack
of sensitive polarization detectors, in part due to the fate of
approved missions and in part because the celestial X-ray sources
appeared less polarized than expected. Only one positive
measurement has been available until now. Indeed the eight
Orbiting Solar Observatory measured the polarization of the Crab
nebula in the 70s. The advent of techniques of microelectronics
allowed for designing a detector based on the photoelectric effect
in gas in an energy range where the optics are efficient in
focusing X-rays. Here we describe the Instrument, which is the
major contribution of the Italian collaboration to the Small
Explorer mission called IXPE, the Imaging X-ray Polarimetry
Explorer, which will be flown in late 2021. The instrument, is
composed of three Detector Units, based on this technique, and a
Detector Service Unit. Three Mirror Modules provided by Marshall
Space Flight Center focus X-rays onto the detectors. In the
following we will show the technological choices, their scientific
motivation and the results from the calibration of the Instrument.
IXPE will perform imaging, timing and energy-resolved polarimetry
in the 2-8 keV energy band opening this window of X-ray astronomy
to tens of celestial sources of almost all classes.

\end{abstract}



\section{Introduction} \label{sec:intro}
Historically, since the dawn of X-ray Astronomy, polarimetry was
considered a 'holy grail' due to the characteristics of the
radiation emitted by the celestial sources and their, typically,
non-spherical geometry. Rockets were flown in the late sixties
\citep{Angel1969} and in the early seventies \citep{Novick1972}
with polarimeters based on the 'classical' technique of Bragg
diffraction \citep{Schnopper1969} and Thomson scattering
\citep{Angel1969}. A rocket experiment in the 1970s hinted at
polarized emission from the Crab Nebula and Pulsar
\citep{Novick1972}. The eight Orbiting Solar Observatory (OSO-8)
X-ray polarimeter confirmed this result with a much higher
significance ($\Pi$ =19.2\%$\mathrm{\pm}$1.0\% @ 2.6 keV,
\cite{Weisskopf1978}) definitively establishing the synchrotron
origin of the nebular X-ray emission. Due to its limited observing
time and relatively high background, the OSO-8 X-ray Bragg
polarimeter obtained useful upper limits for just a few other
bright galactic X-ray sources, while for many others the upper
limits were too coarse to constrain models
\citep{Weisskopf1977,Weisskopf1978b,Hughes1984}. From the advent
of X-ray telescopes starting with the Einstein satellite, it was
clear that a quantum leap required a focal plane instrument. An
experiment based on the classical techniques was devised and built
but never flown\citep{Kaaret1989,Soffitta1998,Tomsick1997}.

An experiment tuned to the classical energy band of X-ray
Astronomy (2-8 keV for the IXPE) allows for meaningful polarimetry
of basically all classes of celestial sources with, maybe, the
exception of clusters of galaxies. To be effective in this energy
band, polarimetry must be based on the photoelectric effect, a
long-sought method by means of which it is also possible to derive
the impact point, the energy and time with suitable devices
\citep{Heitler1954,Sanford1970,Soffitta2001}. \emph{Classical}
methods like Bragg diffraction \citep{Schnopper1969,Novick1972}
are effective, only, in one or two narrow energy bands or, like
Thomson scattering, \citep{Angel1969} suffer of high background
and high energy threshold. The band, the modulation factor (see
Section \ref{sec:statiatics}) and the efficiency of a
photo-emission polarimeter required a complex trade-off in the
choice of the gas mixture, of its pressure, and of the detector
drift length. This was the results of years of extended
simulations and measurements \citep{Muleri2008,Muleri2010a}.

\begin{figure}[ht]
\caption{The Imaging X-ray Polarimetry Explorer with the three
Detector Units at the focus of the three mirror modules.}
\label{fig:IXPE} \centering
\includegraphics[width=5.1in, height=2.83in, keepaspectratio=true]{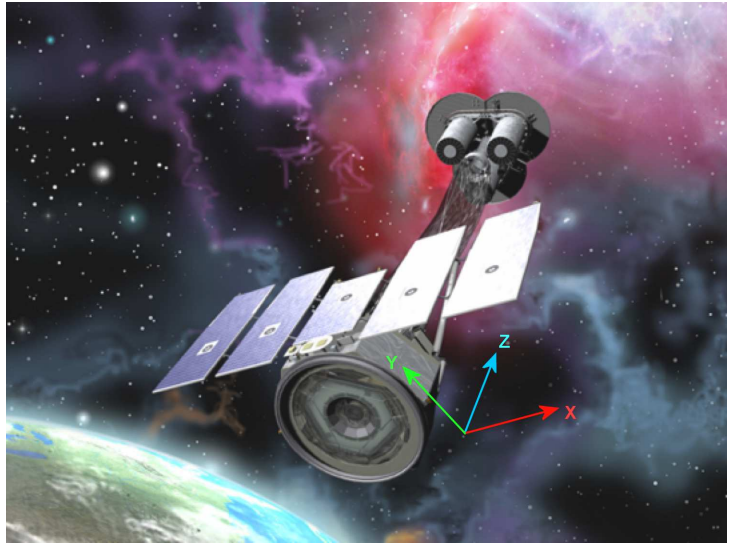}
\end{figure}

The advent of techniques typical of microelectronics allowed to
build sensitive polarimeters with imaging capabilities
\citep{Costa2001,Bellazzini2006,Bellazzini2007}. Later non-imaging
polarimeters, with large quantum efficiency \citep{Black2007}, but
needing rotation due to spurious effects, intrinsic to the readout
methodology, were also devised. The Gravity and Extreme Magnetism
SMEX (Small Explorer) \citep{Swank2010} based on the latter
technology was eventually selected, but then canceled in 2012 for
programmatic reasons. A mission similar to IXPE was proposed as a
small ESA mission \citep{Soffitta2013} and then, redesigned, as
MIDEX ESA mission down-selected for competitive phase A but not
approved for flight \citep{Soffitta2017b}. IXPE
\citep{Weisskopf2008,Weisskopf2016}, see Figure \ref{fig:IXPE}, is
a SMEX scientific mission selected by NASA in January 2017 with a
large contribution of Agenzia Spaziale Italiana (ASI) to be
launched in 2021 aboard a SpaceX Falcon-9 rocket. It will measure
polarization in X-rays from neutron stars, from stellar-mass black
holes and from Active Galactic Nuclei (AGN). For the brightest
extended sources like Pulsar Wind Nebulae (PWNe), Super Nova
Remnants (SNRs), and large scale jets in AGN, IXPE will perform
angularly resolved polarimetry. In some cases, IXPE will
definitively answer important questions about source geometry and
emission mechanisms or even exotic effects in extreme
environments. In others, it will provide useful constraints
including the possibility that no currently proposed model
explains the data.

A Gas Pixel Detector (GPD), based on the same technology exploited
in IXPE \citep{Baldini2021}, was flown on board a CubeSat and is,
presently, collecting scientific data from a long observation of
the Crab Nebula, with a possible detection of a time variation of
polarization of the pulsar \citep{Feng2020}, and other bright
sources.

The sensitivity of IXPE is 30 times better than that of the Bragg
polarimeter on-board OSO-8, for a 10 mCrab source, while is more
then two order of magnitude better then the GPD on board of the
Cubesat for a source of the same intensity. While polarimetry of
Active Galactic Nuclei was precluded to OSO-8, it is within the
reach of IXPE which provides, also, imaging capability in
combination with simultaneous spectral and temporal measurements.
IXPE's design is based on the scientific requirements summarized
in the Table \ref{tab:SciReq}.

A characteristic of IXPE is the dithering during observations.
This has been introduced to avoid having to calibrate pixel by
pixel, as with Chandra and other observatories with imaging
detectors (see Section \ref{sec:GPD}). Modern X-ray telescope
missions are designed for staring or dithering observation Each
detector is mounted, with respect to the next one, on the top-deck
of the spacecraft with an angle, around  the Z-axis, (the axis of
the incoming photon beam, see Figure \ref{fig:IXPE}) of
120$^\circ$. Such disposition allow for a sensitive reduction and
check of spurious effects.

\begin{deluxetable*}{cccc}[ht!]
\tablecaption{IXPE Scientific Requirements. Half Power Diameter
(HPD), Field of View (FOV) and Minimum Detectable Polarization
(MDP) are shown\label{tab:SciReq}} \tablewidth{0pt} \tablehead{
\colhead{Physical} & \colhead{Observable} & \colhead{Property} &
\colhead{Value}\\
\colhead{Parameter}&\colhead{}&\colhead{}&\colhead{}} \startdata
Linear & Degree \textit{$\mathit{\Pi}$}, angle \textit{$\psi$} &
Sensitivity MDP${}_{99}$(F${}_{2-8\ }$=10${}^{-11}$ cgs,
$\mathrm{\Delta}$\textit{t} = 10 d) & $\mathrm{\le}$ 5.5\%\\
Polarization &    & Systematic error in polarization degree
\textit{$\mathit{\Pi}$ }(5.9 keV) & $\mathrm{\le}$ 0.3\%\\
   &          & Systematic error in position angle \textit{$\psi$} (6.4 keV) & $\mathrm{\le}$ 1$\mathrm{{}^\circ}$\\
Energy & \textit{F}(\textit{E}),
\textit{$\mathit{\Pi}$}(\textit{E}),
\textbf{\textit{$\psi$}}(\textit{E})\textit{} & Energy band
\textit{E}${}_{min}$--\textit{E}${}_{max}$ & 2--8 keV\\
dependence &          & Energy resolution $\mathrm{\Delta}$\textit{E }(\textit{E} = 5.9 keV), \textbf{$\propto$}$\mathrm{\sqrt{}}$\textit{E} & $\mathrm{\le}$ 1.5 keV\\
&&&\\
Spatial& \textit{F}(\textit{k}),
\textit{$\mathit{\Pi}$}(\textit{k}),
\textbf{\textit{$\psi$}}(\textit{k})\textit{} & Angular resolution
HPD (system-level) & $\mathrm{\le}$ 30$^{''}$\\
dependence &  & Field of view FOV $\mathrm{>}$$\mathrm{>}$ HPD & $\mathrm{\ge}$ $9^{'}$\\
&&&\\
Time & \textit{F}(\textit{t}),
\textit{$\mathit{\Pi}$}(\textit{t}),
\textbf{\textit{$\psi$}}(\textit{t}) & Time accuracy
$\mathrm{<}$$\mathrm{<}$ source pulse periods & $\mathrm{\le}$
0.25 ms\\
dependence &&&\\
&&&\\
Areal & \textit{R}${}_{B}$/\textit{A}${}_{det}$ &
\textit{R}${}_{B}$/\textit{A}${}_{det}$$\mathrm{<}$$\mathrm{<}$
\textit{R}${}_{S}$/\textit{A}${}_{S}$ for faint source
(2-8 keV, per DU) & $\mathrm{<}$ 0.004 s${}^{-1}$cm${}^{-2}$\\
background &&&\\
rate &&&\\
\enddata
\end{deluxetable*}

\section{The statistics for an experiment of polarimetry}
\label{sec:statiatics}

A detector system can measure polarization if its response is
modulated by the polarization of the incoming photons. The
photoelectric effect modulates the emission direction of an
s-photoelectron as $\cos{^2}$ of the azimuthal angle, which is the
emission angle with respect to the polarization angle. This
$\cos^{2}$ dependence is a general feature for a polarimeter. The
modulation curve is the angular distribution of the emission
directions of photoelectrons in our case and the peak positions
determine the polarization angle. The amplitude of the modulation
is defined as the semi-amplitude of the modulation curves
normalized to its average. If the incoming radiation is 100$\%$
polarized the modulation amplitude is the so called modulation
factor $\mu$. This is a key parameter of a polarimeter and spans
from 0, if the instrument is not sensitive to polarization, to 1,
in case of an ideal polarimeter.

The sensitivity of a polarimeter is expressed in terms of minimum
detectable polarization ($MDP_{CL}$) where CL is the confidence
level:

\begin{equation}
MDP_{CL} = \sqrt{-2ln(1-CL)} \times \frac{
 \sqrt{2 (C_{S}+C_{B})}}{\mu \times C_{S}} = \sqrt{-2ln(1-CL)}
 \times \sigma_{P}
\end{equation}

C$_{S}$ are the total source counts whereas C$_{B}$ are the total
background counts. We usually set CL = 0.99 to get the $MDP_{99}$
and we get:

\begin{equation}
MDP_{99} = 4.29 \times \frac{\sqrt{C_{S}+C_{B}}}{\mu \times C_{S}}
\end{equation}

\begin{equation}
MDA_{99} = 4.29 \times \frac{\sqrt{C_{S}+C_{B}}}{C_{S}}
\end{equation}

The minimum detectable amplitude is the minimum modulated signal
measurable from a polarized source at 99$\%$ confidence level. It
consists of the MDP not normalized by the instrumental modulation
factor $\mu$. The raw measurement, indeed, consists of an
histogram of photoelectron emission directions (the so-called
modulation curve) modulated by the X-ray polarization of the
incoming beam.

If a polarization is detected with a modulation amplitude equal to
$MDA_{99}$, the measurement is only at about the 3-sigma
\citep{Elsner2012} :

\begin{equation}
n_{\sigma} = \sqrt{-2 \times ln(1-CL)} = 3.03
\end{equation}

Sad to say, but X-ray polarimetry is possible only for relatively
bright celestial sources due to the large number of counts needed
to build the modulation curve. For a focal plane polarimeters, the
background in case of an observation of a point-like source
consists, only, of un-rejected events located within the Point
Spread Function of the optics. As a matter of fact, for these two
reasons the background, for all the point-like celestial sources
suitable for polarimetry, is negligible \citep{Xie2021}. The
sensitivity, therefore, depends, in this case inversely with
respect to $\mu$ and as a square root of the total source counts.
In turn the total number of the source counts depends linearly on
the observing time $T$, mirror effective area $A_{eff}$,
efficiency of the detector $\epsilon$ and source spectrum $S$. In
this case the Minimum Detectable Polarization (MDP) can be
expressed as:

\begin{equation}
\label{eq:MDP&QF}
 MDP_{99} = \frac{4.29}{\mu \sqrt{S A_{eff} \epsilon T}}
\end{equation}

The quality factor of the detector can be expressed as QF = $\mu$
$\times$ $\sqrt{\epsilon}$. A trade-off between $\mu$ and
$\epsilon$ provides the best sensitivity to polarization
obtainable by the detector.

We also use the Stokes parameters to characterize the
polarization. These have the advantage that, as opposed to the
degree of polarization and the position angle, they are
statistically independent. According to the prescription in
\citep{Kislat2015} we calculated the Stokes parameters of each
count. We sum-up the corresponding Stokes parameter to determine
the polarization degree and polarization angle of the beam. By
fitting a cosine function or summing the Stokes parameter, we
obtain the same results.

\section{The scientific drivers of the Instrument}
IXPE scientific requirements descend from the expected
polarization properties of the celestial sources and their
temporal, spectral and spatial characteristics. The sensitivity of
IXPE, as shown in Table \ref{tab:SciReq}, is set for meaningful
polarimetry, with realistic observing time, of the brightest AGNs.
The residual systematic error in the modulation amplitude should
allow for polarimetry below 1$\%$ when observing bright galactic
sources. The requirement on the systematics of the polarization
angle allows to check possible variation with respect to the old
OSO-8 polarimetry of the Crab Nebula. Also, it allows for an
effective comparison, at different wavelengths, of phase resolved
X-ray polarimetry in combination with the requirement on the
timing accuracy. The requirement on angular resolution is derived
by the capability to angularly separate the central pulsar from
the surrounding torus in the Crab Pulsar Wind Nebula.

\begin{deluxetable*}{cc}
\tablecaption{IXPE Instrument Requirements. Where indicated the
20$\%$ cuts refer to an optimum selection of the data (see for
example Section \ref{sec:statiatics}) that maximizes the
sensitivity considering the events with the most elongated tracks.
\label{tab:InstReq}\label{tab:InstReq}} \tablewidth{0pt}
\tablehead{ \colhead{Parameter} & \colhead{Requirement}}
\startdata
Modulation factor & $>$ 27.7 at 2.6 keV (with 20 $\%$ cut)\\
Modulation factor & $>$ 53.7 at 6.4 keV (with 20 $\%$ cut)\\
Spurious modulation & $<$ 0.27 $\%$ at 5.9 keV\\
GPD quantum efficiency & $>$ 17.7 $\%$ at 2.6 keV \\
GPD quantum efficiency &$>$ 1.8 $\%$ at 6.4 keV \\
Energy resolution & 1.5 keV (at 5.9 keV)\\
Knowledge of the spurious modulation & $<$ 0.1 $\%$ at 2-8 keV
keV\\
Systematic error on angle & $<$ 0.4 deg\\
Position resolution (HPD) & $<$ 190 $\mu$m at 2.3 keV \\
Dead time & $<$ 1.2 ms (average at 3 keV)\\
Maximum counting rate & 900 c/s\\
Time accuracy & $\pm$ 94 $\mu$s (99 $\%$) \\
Background & $<$ 0.004 s$^{-1}$ cm$^{-2}$ per DU (2-8 keV)\\
\enddata
\end{deluxetable*}

Table \ref{tab:InstReq} flows down the scientific requirements
into instrument requirements. As a matter of fact, polarization
sensitivity depends inversely on the modulation factor and
inversely to the square-root of the quantum efficiency. The
requirement on the modulation factor is more stringent with
respect to the quantum efficiency. As a matter of fact a deficit
in quantum efficiency results in a linear increase of the
observing time to get the same sensitivity. A deficit in the
modulation factor requires, instead, a quadratic increment.

The requirement on the maximum allowed modulation amplitude from
an unpolarized source was set, initially, only at 5.9 keV because
of the availability of a true unpolarized source. A low energy
effect (see Section \ref{sec:GPD}) was found and characterized.

Time resolution and time accuracy requirement are set accordingly
to what the technology offers thanks to the use of a Global
Positioning System (GPS) receiver, therefore with a large
discovery space in case of pulsating neutron stars either isolated
or in binary systems.

The largest contribution to the angular resolution of the
telescope system is the HPD of the mirror \citep{Fabiani2014}. The
second contribution is the inclined penetration of X-rays onto the
detector which has an active volume with 1cm-deep thickness. The
third contribution is the position resolution of the detectors
which is negligible with respects to the other two. The
requirement on the blurring (position resolution) of the detector
of Table \ref{tab:InstReq} is set not to spoil the angular
resolution of the telescope by more that few $\%$.

The detector has been designed with sufficient FOV when coupled to
mirrors having focal length of few meters. With such focal length
mirrors are efficient for reflecting X-rays in the classical 1-10
keV energy band. The FOV requirement allows for polarimetry with a
single observation of shell-like Supernovae like Cas A, Tycho and
Kepler \citep{Soffitta2013b}.

The energy resolution requirement allows for studying
energy-dependent phenomena, like, for example, the energy
dependent rotation of the polarization angle in galactic
black-holes binaries in the thermal state
\citep{Stark1977,Connors1977}. Moreover this requirement allows
for assigning the correct modulation factor, rapidly increasing
with energy, to the detected photons deriving the polarization of
the X-ray sources.

A small dead-time allows for a minimum decrement of the detected
counting rate, and therefore of the sensitivity, from sources as
bright as the Crab Nebula. Due to the limitation of the onboard
memory and the requirement on the use of the S-band for on-ground
data transmission, particular care must be taken to interleave
bright source observations with faint source observations.

The background-rate requirement, albeit larger than that of space
experiments with gas detectors with front and rear
anti-coincidence, allows for measuring polarization from dim and
extended sources like molecular clouds in the galactic center
region.

\section{IXPE payload overview}
The IXPE payload \citep{ODell2018,ODell2019,SOffitta2020} consists
of a set of three identical telescope systems co-aligned to the
pointing axis of the spacecraft and with the Star-Trackers. Each
system, while operating independently, comprises of a 4-m-focal
length Mirror Module Assembly (MMA, \cite{Ramsey2019}) that
focuses X-rays onto the respective Detector Units (DUs). Each
Detector Unit (DU) hosts one polarization-sensitive imaging
detector, with its own electronics, which communicates with a
Detector Service Unit (DSU) interfaced to the spacecraft
Integrated Avionics Unit (IAU). The three DUs and the DSU are
collectively called the IXPE Instrument. The DUs are mechanically
mounted onto the top deck of the spacecraft oriented, as already
mentioned above, with a rotation of 120$^\circ$ each with respect
to the beam axis (Z-axis see Figure \ref{fig:IXPE}).

The performance of the detectors and, particularly, the gain, have
to be monitored during the operative life of the mission. Other
parameters like the modulation factor, are expected, from ground
calibration, to be more stable in time but they can be monitored
as well. For these reasons each DU is equipped with a
multi-function filter and calibration wheel assembly that allows
for (i) in-flight calibration of the modulation factor,(ii) the
calibration of the gain, (iii) source flux attenuation (iv)
background measurements. MMA and DUs are separated by an able to
be deployed (3.5 m) boom; the position of the MMA with respect to
the DUs can be adjusted after deployment with a Tip/Tilt/Rotate
mechanism. Each MMA hosts an X-ray shield to avoid, in combination
with a stray-light collimator mounted onto the top of the DU,
X-ray photons impinging on the detector active area when arriving
from outside the telescope field of view. The use of a set of
telescopes provides many advantages with respect to a
configuration based on a single telescope of equivalent collecting
area. First of all, since the energy band-pass fixes the ratio
between the mirror diameter and the focal length, a multiple
telescope configuration is more compact than a single larger
telescope; this occurs at the cost of an increase of the measured
background which, however, is not a driving requirement for IXPE.
Moreover, a multiple system is intrinsically redundant and offers
the possibility of comparing independent data and to correct small
effects which may mimic a real signal.

\section{The design of the instrument}
\label{sec:InstrumentProductTree} The design of the instrument,
and in particular the design of the DUs, is the consequence of the
prescription of using the Pegasus XL launcher at the time of the
proposal. The top-deck of the spacecraft hosts at launch both the
three un-deployed mirrors and the three DUs. Hence, the available
space in the top deck requires a vertically stacked mechanical
configuration for the DUs. Such configuration is demanding in
terms of thermal dissipation and, in addition, has a worse
response to vibrations than a side-by-side configuration. In the
following phase of competitive bidding during phase B, a Falcon-9
was eventually selected and will lift IXPE to the foreseen 600 km
equatorial orbit.

\begin{figure}[ht]
\caption{Instrument Product Tree.} \label{fig:ProdTree} \centering
\includegraphics[width=7.07in, height=4.0in, keepaspectratio=true]{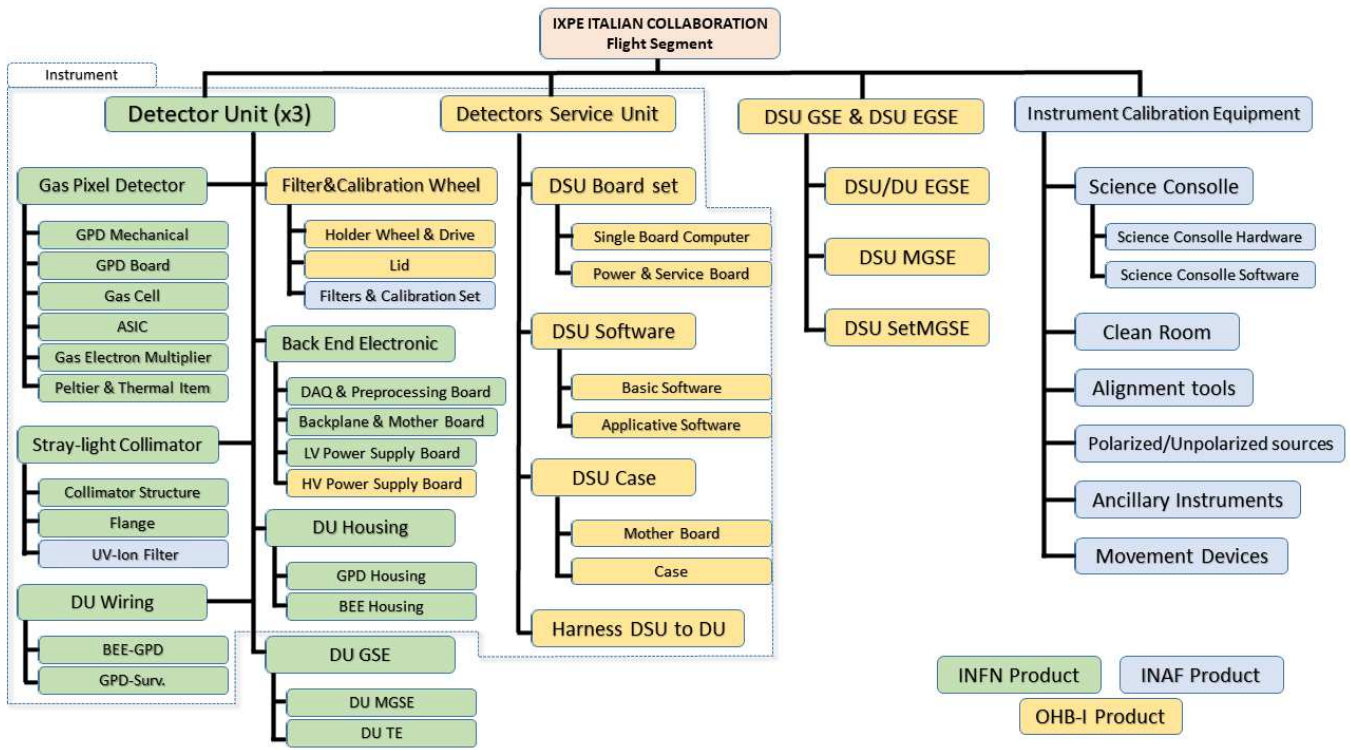}
\end{figure}

The IXPE Instrument, mainly supported by the Italian Space Agency,
ASI, has been entirely designed, built and tested in Italy. We
show the items of the Instrument as Italian IXPE contribution in
Figure \ref{fig:ProdTree}. In the same figure the responsibilities
for each item that are shared between INAF, INFN and the
industrial partner OHB-Italy are shown.

Each of the four DUs (including one spare unit, see Figure
\ref{fig:DU}), comprises the following sub-units:

\begin{enumerate}
\item  \textbf{Gas Pixel Detector} (GPD), which is an X-ray
detector with gas as the absorption medium and a
Application-Specific Integrated Circuit (ASIC) as the readout
electrode, specifically developed by INFN in collaboration with
INAF-IAPS for X-ray polarimetry.

\item \textbf{Filter \& Calibration Wheel} (FCW), which hosts the
calibration set comprising calibration sources and filters for
specific observations to be placed in front of the GPD when
needed.

\item \textbf{Back End Electronics} (BEE), which comprises of
electronics boards (DAQ, Data Acquisition Board,
\cite{Barbanera2021} to manage the GPD ASIC), the required High
Voltage lines and the Low Voltages lines.

\item \textbf{Stray-light Collimator} (STC), already mentioned
above.

\item \textbf{DU Housing} (DUH), which provides the mechanical and
thermal interface of the DU to the S/C.

\item \textbf{DU wiring} (DUW), which provides the electrical
interfaces (internal to the DU) between the BEE and the GPD.
\end{enumerate}

The DSU is the unit which provides the DU with the needed
secondary power lines, controls and powers the FCW, formats and
forwards the scientific data of the three DUs to the spacecraft.
The DSU comprises the following sub-units:

\begin{enumerate}
\item \textbf{DSU Board Set} (DBS), which is the set of electronic
boards (both nominal and redundant) which perform the DSU tasks;

\item \textbf{DSU Software}, which comprises the software which
runs in the DSU;

\item  \textbf{DSU Case} (DSC), which includes a back-plane that
provides the electrical interface among the DSU boards. Further,
the DSU Case provides the mechanical and thermal interface of the
unit;

\item  \textbf{Harness DSU to DU}, which comprises the cables
necessary to electrically interface the three DUs to the DSU.
\end{enumerate}

Moreover, to complete the instrument related units, two
testing/calibration stations were assembled. One is the Instrument
Calibration Equipment, which is shortly described in Section
\ref{sec:ICE}, and the companion Assembly Verification and Test
Equipment. The latter has three positions, one for each DU. Each
testing/calibration stations consist of X-ray sources,
collimators, crystals and computer controlled stages to calibrate
the DUs and to perform the bench integration of the instrument.
The Electrical Ground Support Equipment has also been assembled to
manage the DUs and DSU calibration activities and bench
integration.

In this paper we present the status of the instrument before the
assembly with the spacecraft. However, due to its design, any
change of performances after the integration is not expected.

\section{The IXPE Detector Unit}
\label{sec:DU}
\begin{figure}
\caption{The IXPE Detector Units (a) Exploded view of a Detector
Unit, (b) Photograph of a Flight Detector Unit in its handling
plane. The thermal environment is controlled by means of a Peltier
cooler and a heater. A thermal strap conducts the heat from the
GPD Peltier/heater system to the radiator of the spacecraft.}
\label{fig:DU}
          \gridline{\fig{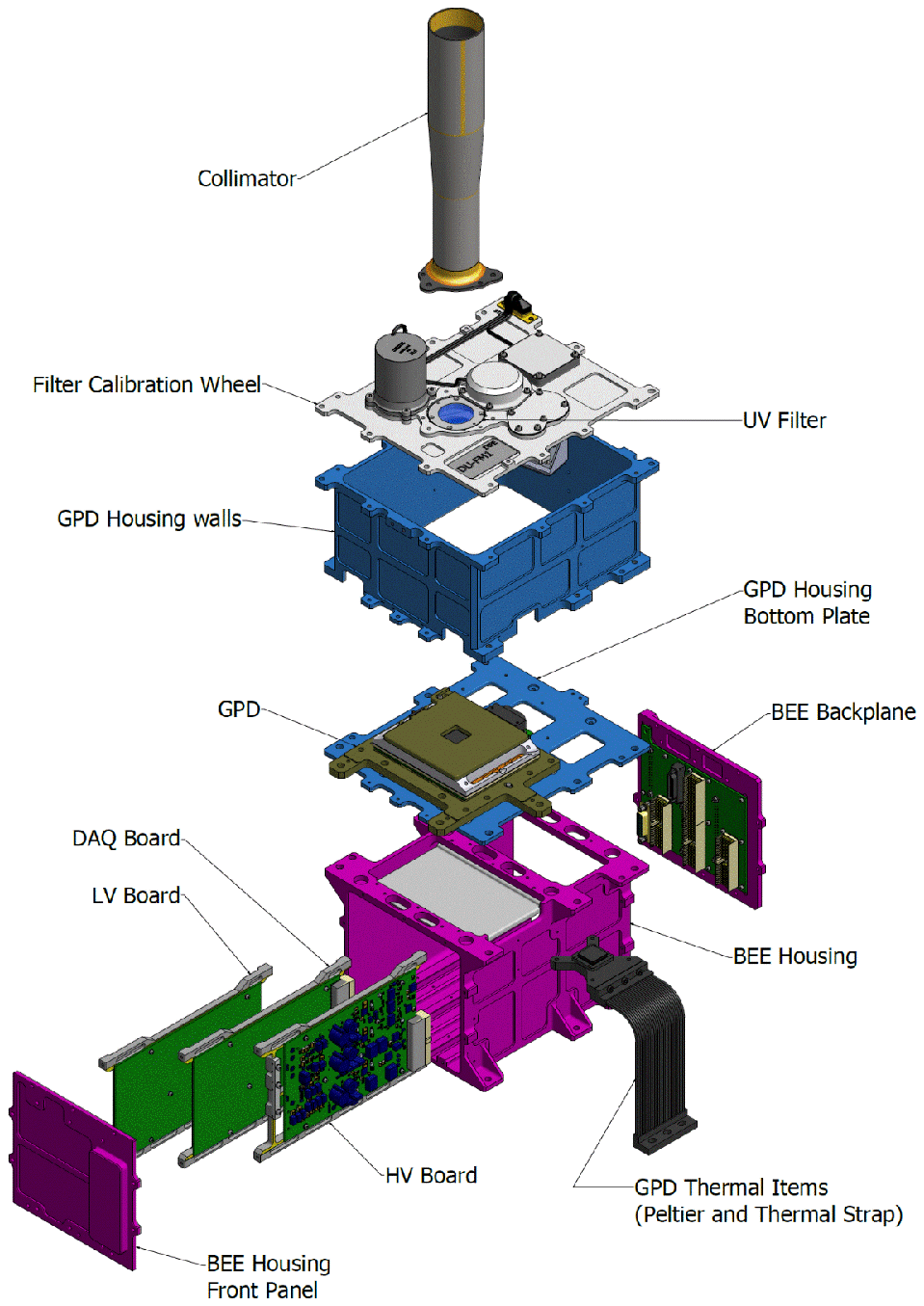}{0.6\textwidth}{(a)}
          \fig{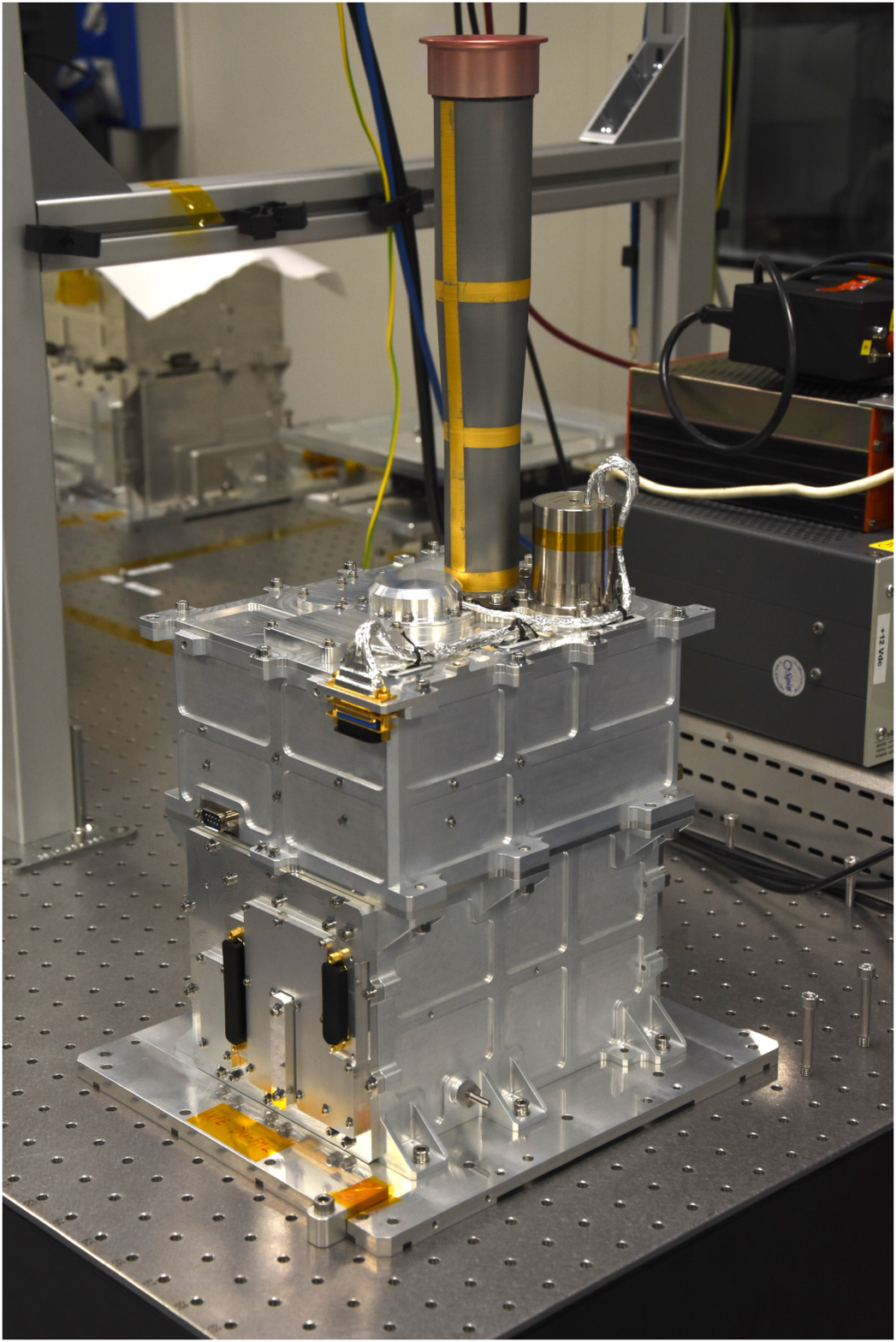}{0.4\textwidth}{(b)}
          }
\end{figure}

In order to save space, the DUs were designed with a
top-bottom configuration for the detector and associated
electronics respectively. The 120$^\circ$ clocking allows for
checking the level of absence of spurious modulation and limiting
any systematic effects. The exploded view of the DU and a picture
of one assembled DU is shown in Figure \ref{fig:DU}.

The DU housing is composed of two boxes. The top one is the "GPD
housing". The bottom item is the "Back-End Housing". In the next
sections we describe the main elements of the DUs.

\subsection{Stray-light collimator and UV-ion filter}
\label{sec:SLC&UV-ionFilter} An extensible boom connects the
mirror modules to the spacecraft top-deck. This open-sky
configuration requires a system to prevent that Cosmic X-ray
Background photons arriving from outside the mirror field of view
impinge on the detector. At this aim, a tapered stray-light
carbon-fiber (CF) collimator collects only the photons reflected
by the optics, thanks, also, to fixed X-ray shields around the
mirror modules. The thickness of the collimator is 1.25 mm and
includes the 1-mm thick CF with an external 50-$\mu$m molybdenum
coating and 20 $\mu$m gold of the molybdenum foil.

Due to the fact that the beryllium window of the detector and the
supporting titanium frame is at high potential (about -2800 V)
positive ions may interact with the top GPD structure producing
eventual secondary photons and a possible failure of the high
voltage system. For this reason, photons from the optics cross a
UV-ion filter made by LUXEL composed of 1059 nm of
LuxFilm$\textsuperscript{\textregistered}$ (based on kapton) with
an external coating of 50 nm of aluminum and an internal coating
of 5 nm of Carbon. The UV-ions filters are described in
\cite{LaMonaca2021}. The scope of UV-ions filter is to prevent UVs
and low velocity plasma present in orbit for reaching the
beryllium window of the GPD (see Section \ref{sec:GPD}). The
transparencies at two energies are reported in Table
\ref{tab:InstrumCharact}.

\subsection{The Gas Pixel Detector in the Detector Unit}
\label{sec:GPD} 
\begin{figure}[ht]
\caption{Schematic representation of GPD assembly.}
\label{fig:GPDExpView} \centering
\includegraphics[width=2.2in, height=3.46in]{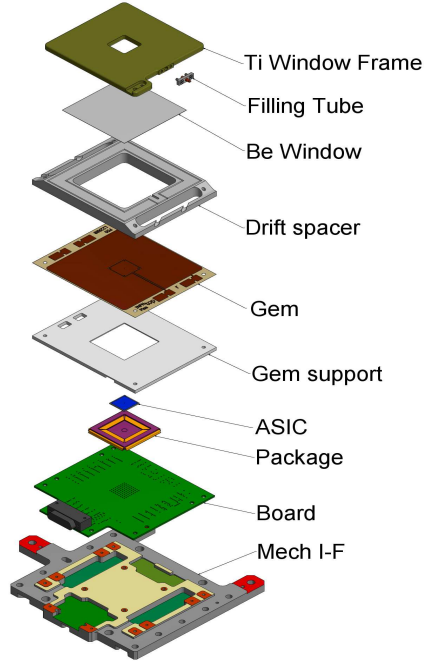}
\end{figure}

\begin{figure}[ht]
\caption{Real photoelectron track at 5.9 keV with reconstructed
direction of emission and absorption point.} \label{fig:track}
\centering
\includegraphics[width=6.04in, height=3.83in,keepaspectratio=true]{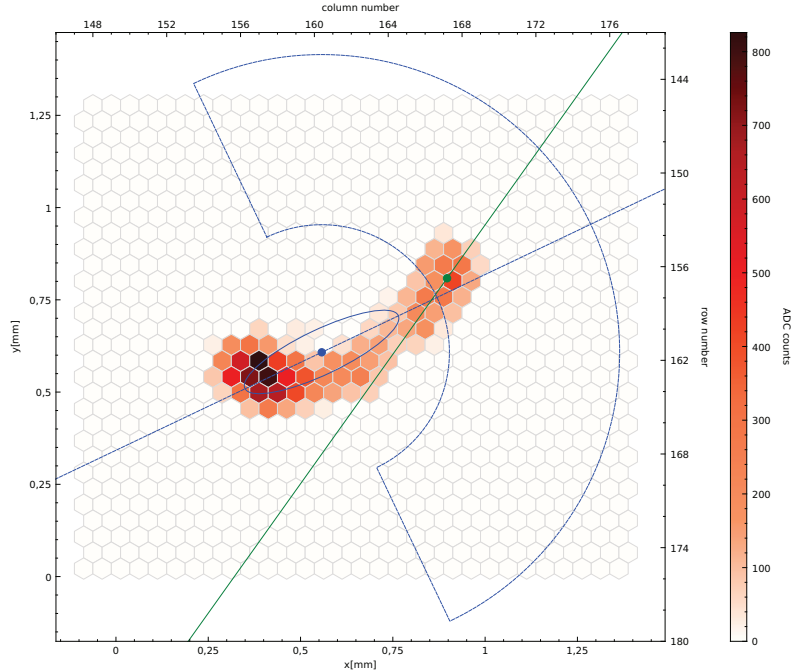}
\end{figure}

The GPD (see Figure \ref{fig:GPDExpView}), located
in the GPD housing, is a gas detector for which the charge-signal
is readout by a matrix of 105k pixels (300 $\times $352 pixels
arranged in a 50 $\mu$m pitch hexagonal pattern) of a dedicated,
custom Complementary Metal-Oxide Semiconductor (CMOS) ASIC
\citep{Bellazzini2006}. The custom ASIC has self-triggering
capabilities thanks to local triggers defining each group of four
pixels called mini-clusters. Each event consists of a Region of
Interest (ROI) made by all the mini-clusters that trigger plus a
selectable additional fiducial region of 10/20 pixels. The charge
content of each pixel in the ROI is readout serially from a single
buffer as differential current output by means of a 5 MHz clock.
The ASIC, also, provides the absolute position of the ROI as
digital coordinates of two opposite vertices and the global
trigger output, about 1$\mu$s after the arrival of the charge. The
threshold is set low enough to collect signals coming from events
which release about 300 eV. The charge amplification is provided
by a Gas Electron Multiplier (GEM), a thin 50 $\mu$m dielectric
(liquid crystal polymer) foil with 9-$\mu$m copper metallization
on both sides. Through-holes are laser etched and disposed on a
regular triangular pattern with vertexes 50$\mu$m apart. These
cylindrical holes have a diameter of 30 $\mu$m
\citep{Tamagawa2009,Tamagawa2010}. The basic building blocks of
the GPD assembly and their functions are schematically illustrated
in Figure \ref{fig:GPDExpView}. The small pitch of the ASIC and
the GEM are responsible for the good image of the photoelectron
track (see Figure \ref{fig:track}).

In summary, the GPD is composed of the following subassemblies:

\begin{enumerate}
\item  \textbf{mechanical interface} (\textit{GPD Mech. I/F}),
made of titanium, which supports the GPD unit, connects it to the
focal plane and provides references for alignment with the MMA;
\item  \textbf{printed circuit board} (\textit{GPD Board}), which
connects the GPD to the readout electronics; \item \textbf{ceramic
spacer} (\textit{GEM Support Frame}), supporting and insulating
the GEM from the GPD Board; the lower gas gap of 770 $\mu$m for
the drift of the amplified electrons is defined by the ASIC top
surface and the bottom GEM. \item the \textbf{GEM foil}
(\textit{GEM}), including four soldering pads for the high voltage
connections; \item \textbf{ceramic support} (\textit{Drift
Spacer}), which defines the absorption gas cell above the GEM and
isolates the GEM from the top electrode; \item \textbf{titanium
frame} (\textit{Ti frame and Be window}) which closes the gas cell
and allows X-rays into the GPD through the integrated thin,
optical-grade 50 $\mu$m one-side aluminized beryllium window. The
titanium frame and the beryllium window serve, also, as a drift
electrode; \item \textbf{filling tube} of Oxygen Free High
Conductivity copper and its fixture to the titanium frame
(\textit{tube and fixture});
\end{enumerate}

The GPD is sealed and does not require any gas cycling systems.
The keystone of the assembly is the Kyocera package. In fact the
ASIC is designed to fit into a commercial package that acts as the
bottom parts of the gas cell and connects the internal wiring and
the clean gas volume to the external Printed Circuit Board (PCB),
onto which the package is soldered. In this way cleanness is
preserved in the gas volume.

The GPD assembly is done at INFN, whereas the filling procedure is
performed at Oxford Instruments Technologies  Oy (OIT, Espoo,
Finland). After a vacuum leak-tightness check, the GPD is
continuously baked-out and pumped-down for 14 days at 100
$^\circ$C for outgassing and eliminating possible impurities in
the internal gas cell. Finally the GPD is filled with a
Dimethyl-Ether (DME) mixture at 800 mbar at room temperature. OIT
filled with the same gas mixture a single-wire proportional
connected in parallel to the same filling station to check the
energy resolution. If the energy resolution is within the
requirement, the filling tube of the GPD is crimped. We selected
this three-times distilled gas mixture because it minimizes track
blurring thanks to its small transverse diffusion ($\frac{69 \mu
\textrm{m}}{\sqrt{\textrm{cm}}}$ at 1833 V/cm and 0.8 bar).  Its
pressure is optimized in order to provide the best combination of
efficiency and modulation factor and, therefore, the sensitivity
in the energy range of interest. Indeed a lower pressure increases
the track length, but also the diffusion and reduces the quantum
efficiency. A larger pressure, instead, increases the quantum
efficiency but reduces the track length and this reduction is not
compensated in terms of modulation factor by the simultaneous
reduction of diffusion. In Section \ref{sec:gain} we will describe
a peculiar phenomenon called 'virtual leak' that reduces
asymptotically the pressure at 640-650 millibar increasing the
track length.

Three negative high voltages are required by the GPD. The GEM
bottom voltage (V${}_{bottom}$ on the side facing the ASIC) sets
the electric field in the transfer gap between the GEM and the
ASIC. The difference between the GEM top voltages (V${}_{top}$ on
the side facing the window) and V${}_{bottom}$ sets the gas gain.
The high voltage on the titanium frame and beryllium window,
(V${}_{drift}$) sets the drift voltage in the absorption region in
combination with the V${}_{top}$.

The performance of the GPD is, in principle, dependent on
temperature to some extent. The GPD temperature is controlled
during the operation by means of a Peltier cooler and heaters. A
thermal strap connects the cooler directly to the spacecraft
radiator. The GPD thermal control is performed directly by the DSU
and is designed to be as independent as possible from the BEE
temperature, since the latter has much looser requirements. The
Peltier cooler and the heaters, and the thermal dissipation of the
spacecraft, maintain the temperature of the GPD within the range
15-30 $^\circ$C.

\section{The physical quantities provided by the GPD}
\label{sec:GPDPhysicalQuantities}

Thanks to the GPD, all the information contained in the observed
X-ray radiation is provided to the user, after that a proper
algorithm is applied to the data (see Section \ref{DataFormat}).
The projection of the track onto the hexagonally patterned top
layer of the ASIC is analyzed to reconstruct the point of impact
and the original direction of the photoelectron. The impact point,
not the charge barycenter, gives an imaging resolution which is
better than 190 $\mu$m in diameter \citep{Soffitta2012}. The
eventual spatial resolution of the instrument is due to the
diffusion of the initial track, to the reconstruction algorithm
capabilities and to the quantization of the GPD pixel size.

The energy is estimated by summing the charge content of each
pixel.

The event time is determined by the trigger output of the ASIC.
Compared with a Charge-Coupled Device (CCD) currently flying in
most of the space mission with X-rays optics, no pile-up issue is
present because of the fast drift time (1 cm/$\mu$s) coupled with
the fast inhibition to additional charge processed by the ASIC.

The linear polarization is determined by the angular distribution
of the emission directions of the photoelectron tracks in the
selected energy band. This is called the modulation curve. If the
radiation is polarized, the modulation curve shows a cosine square
modulation, whose amplitude is proportional to the degree of
polarization and whose phase coincides with the angle of
polarization. The amplitude of the modulation of the incident beam
is eventually normalized to the modulation obtained for 100 $\%$
completely polarized photons (the so called modulation factor
$\muup$) to get the beam polarization.  The energy dependant
modulation factor of the GPD is measured at different discrete
energies during the calibration campaign of the detector and well
agrees with predictions derived by accurate Monte Carlo
simulations.

During the construction phase of the instrument, and in particular
of the Gas Pixel Detector, we detected a non-zero modulation from
unpolarized radiation. This signal is called spurious modulation.
The amplitude of this signal is of the order of 1.5 $\%$, on each
DU, at 2 keV rapidly declining with energy. The root cause of this
spurious modulation is not totally understood yet, but we have
identified a few effects that contribute to it by systematically
deforming the photoelectron track. Some of them, collectively
called ``ASIC effects'', originate from small systematic biases in
the ROI reading. These are typically tiny, i.e., a fraction of the
pixel electronic noise in amplitude, yet they cause a measurable
effect on large number of counts. They were minimized by an
appropriate choice of the read-out clock frequency and eventually
canceled out by subtracting the residual measured bias from the
data in the off-line analysis. Another contribution to spurious
modulation comes possibly from the deformation of the track during
multiplication, as is suggested by the correlation between the
gain spatial mapping and the spurious modulation spatial mapping.
In order to subtract the spurious modulation from the data we
organized a vigourous calibration campaign for each DU in such a
way to dedicate most of the time to characterize this effect. We
used several nearly unpolarized sources at different energies and
on different regions of the detector, from 2 keV up to 5.9 keV
where this effect vanishes. More details on the calibration
campaign are reported in \ref{sec:ICE}.

\subsection{Short $\&$ secular gain variation}
\label{sec:gain}

A short-term rate-dependent gain decreasing, accompanied by a
secular gain increasing, was also detected. The first one is due
to charging effects on the exposed dielectric by drifting ions and
electrons during the multiplication process. The second is likely
due to an absorption or adsorption of the gas by materials inside
the GPD. These two effects have been studied in detail. The first
effect is described by simple functions applied by the processing
pipeline to the data for gain normalization. The second effect is
expected to saturate at an average 640-650 mbar already by the
time of the flight as indicated by the gain, the counting rate
from a $^{55}Fe$ reference source and the track length trends. All
these parameters are, simultaneously, well modelled in ten
detectors including the three flight DUs, the spare DU and the GPD
control sample \citep{Baldini2021}. The impact is a decrement of
the quantum efficiency, however the accompanying increase of the
modulation factor reduces the loss of sensitivity to an acceptable
level.

\subsection{The Filters and Calibration Wheel}
The filter and the calibration wheel (FCW) has been designed to
monitor the performance of the detector during the life of the
mission. In particular we intend to check the low and high energy
modulation factor and the spurious modulation during the
observational life of the mission. Also we monitor the gain that
depends both on temperature and on charging (see Section
\ref{sec:GPD} and \ref{sec:gain}).

 The FCW hosts filters to perform special observations (e.g., very bright
sources) and calibration sources to monitor detector performance
during flight. The FCW has 7 positions commanded by the DSU. The
positions correspond to open position, closed position, gray
filter and calibration source A, B, C and D. The stepper motor is
a Phytron phySPACE$^{TM}$ 42-2. It is a COTS component with 25
years of space flight heritage. The phySPACE$^{TM}$ series is
developed and built to resist vacuum, vibrations, low/high
temperature and radiation while maintaining high performance,
precise positioning and long life. The motor pinion is made of
Vespel$\textsuperscript{\textregistered}$. The double bearing
system that connects the fixed part to the rotating one is made of
stainless steel. The positioning is performed and monitored by two
different systems.

One system is based on the use of Hall sensors as non-contact
switches to stop the wheel rotation at the requested position. A
combination of three hall sensors allows for setting and
monitoring the 7 wheel positions.

In order to have an independent measurement of the wheel position
a second, analog, device, a Novotechnik PRS65/S152 potentiometer,
is implemented. A qualification campaign allowed for
declaring this component as flight compatible.

The FCW positioning can be set digitally or on' a step-by-step
basis permitting gain and modulation factor to be measured at
various positions on the detector.

\begin{figure*}[h]
\caption{CAD models of the calibration set. All sources are
derived from capsules containing an $^{55}Fe$ nuclide: (a) Cal A
is the polarized source. It is un-collimated and provides
simultaneous polarized X-rays at 3.0 keV and 5.9 keV, (b) Cal B is
an unpolarized collimated 5.9 keV source, (c) Cal C is a broad 5.9
keV source for flat field illumination, (d) Cal D is a broad 1.7
keV source for gain and spurious modulation check-out
\label{fig:CalSources}}
\gridline{\fig{CalA.eps}{0.3\textwidth}{(a)}
          \fig{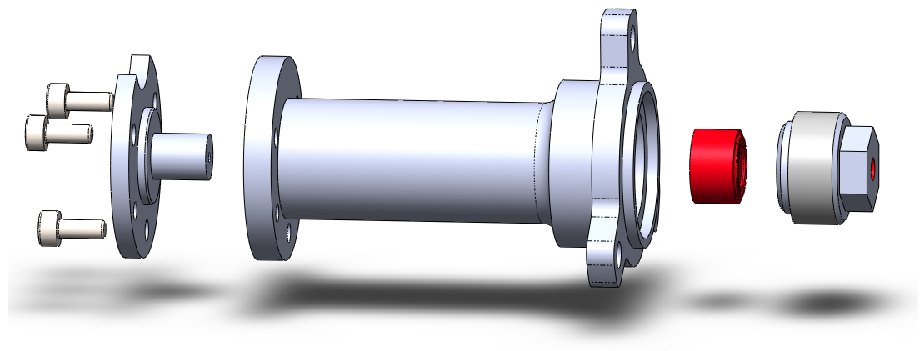}{0.4\textwidth}{(b)}
          }
\gridline{\fig{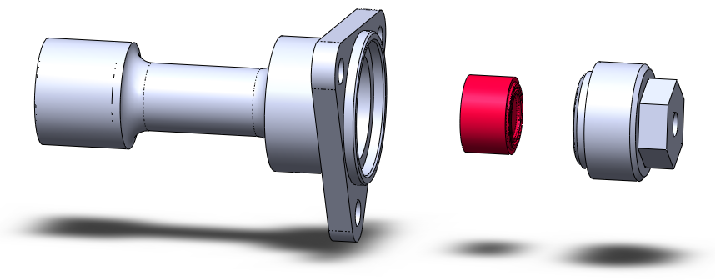}{0.4\textwidth}{(c)}
          \fig{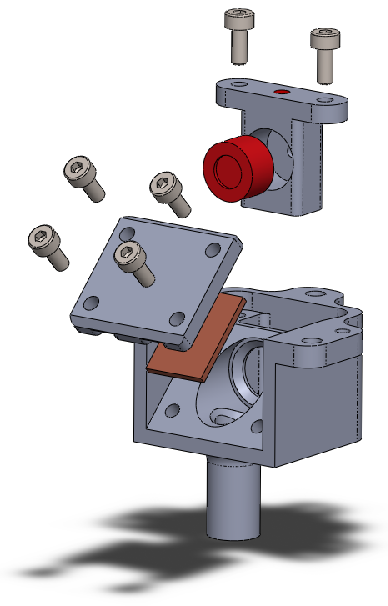}{0.4\textwidth}{(d)}
          }
\end{figure*}

\label{sec:FCS} The calibration set is composed of the X-ray
sources and their holders. It comprises the items which can be put
in front of the GPD by rotating the FCW. The design of the
calibration set is shown in Figure \ref{fig:CalSources}.

\begin{figure}[htpb]
\caption{The measured transparency of the gray filter used in each
DU.} \label{fig:GreyFilterTransp}
\centering
\includegraphics[width=5.07in, height=2.83in,keepaspectratio=true]{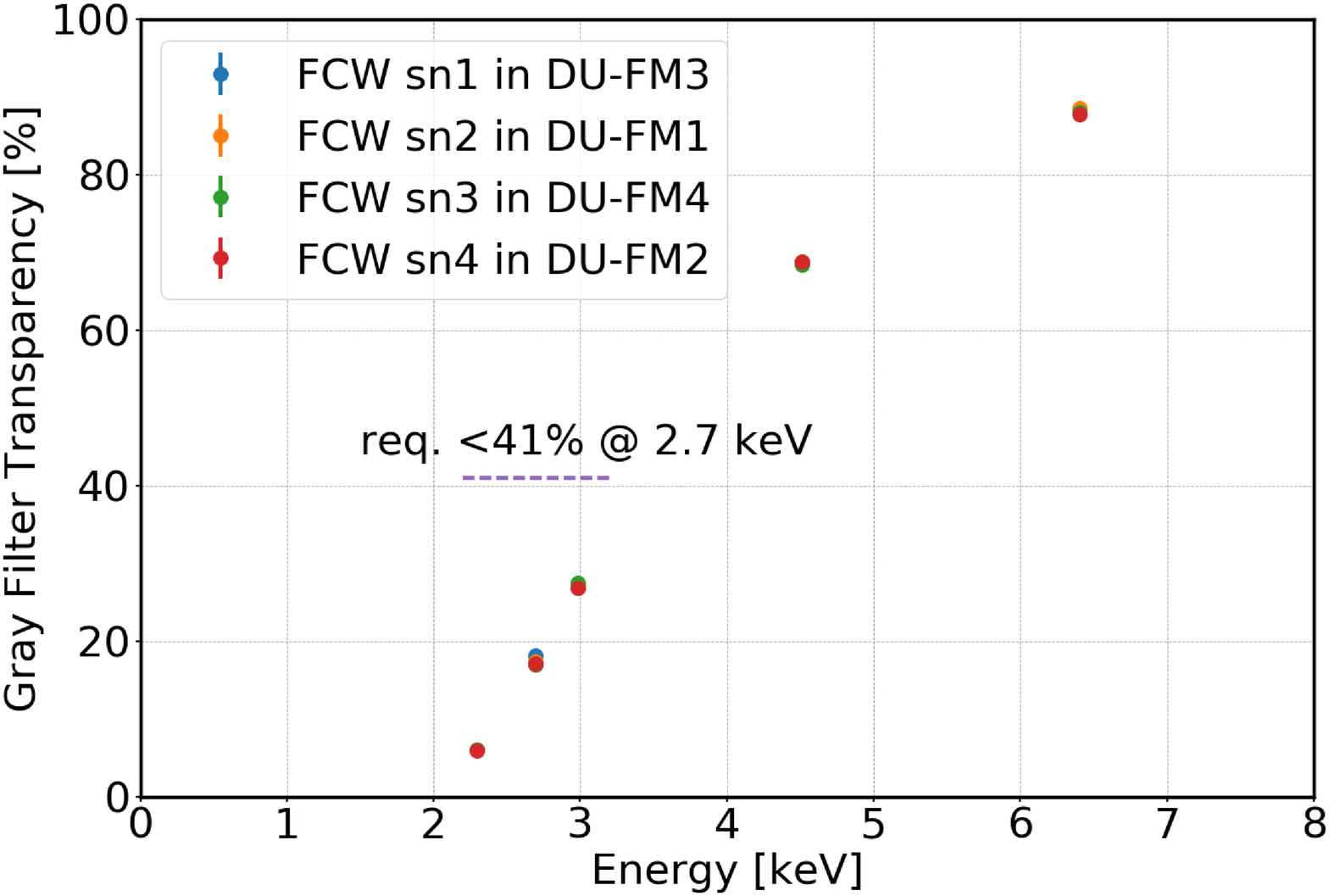}
\end{figure}

In the following we summarize the particular positions which are
also described in \cite{Ferrazzoli2020}:

\begin{enumerate}
\item \textbf{Open position}. The open position is the standard
one used for astrophysical observations. \item\textbf{Closed
position}. The closed position has two motivations. It is a safe
position for the instrument, but it is also necessary to gather
instrumental (internal) background to be compared with that
obtainable during Earth occultation. \item \textbf{Gray filter}.
Because of the relative large, ($>$ 1 msec) dead-time, of the Gas
Pixel Detector and the available throughput from the bus and from
the down-link, we decided to include a partially opaque absorber.
This partially opaque absorber is designed to be used when the
count-rate is greater than that produced by a source about 4 times
more intense than the Crab or greater. It absorbs efficiently
low-energy photons where the count-rates are greater. At higher
energies, where the count-rates are smaller the absorption is
small. The filter is made of 75-$\mu$m thick kapton foil with 100
nm aluminum on both sides. For a Crab-like spectrum (power law
with index 2) the flux will be reduced of a factor about 6 in the
1-12 keV energy band. The measured energy dependent transparency
is shown in Figure \ref{fig:GreyFilterTransp}.
\item\textbf{Calibration source A (Cal A)}. This source produces
polarized X-ray photons at two energies (3.0 keV and 5.9 keV) to
monitor the modulation factor of the instrument in the IXPE energy
band. An exploded view of this source is shown in Figure
\ref{fig:CalSources}. A single ${}^{55}$Fe nuclide is mounted and
glued into a \emph{T}-shaped holder. X-rays from ${}^{55}$Fe at
5.9 keV and 6.5 keV are partially absorbed by a thin silver foil
mounted in front of the ${}^{55}$Fe nuclide to produce
fluorescence at 3.0 keV and 3.15 keV. The silver foil is 1.6
$\muup$m thick and is deposited between two polyimide foils which
are 8 $\muup$m (on the side towards the ${}^{55}$Fe) and 2
$\muup$m (on the opposite side). Photons at 3.0 keV and 5.9 keV,
broadly collimated, are diffracted by a graphite mosaic crystal,
with FWHM mosaicity of 1.2 deg, at first and second order of
diffraction, approximately at the same diffraction angle (about
0.5$^\circ$). A diaphragm is used to block the stray-light X-rays
except the scattered X-rays from the graphite. \item
\textbf{Calibration source B (Cal B)}. This source produces a
collimated beam of unpolarized photons to monitor the
 degree of spurious modulation. An ${}^{55}$Fe radioactive source
is glued in a holder and screwed in a cylindrical body. A
diaphragm with an aperture of 1 mm collimates  X-rays from the
${}^{55}$Fe to produce a spot of about 3 mm on the GPD; such a
spot has a size which is representative of the source image of a
point-like source when the spacecraft dithering  is included.
\item \textbf{Calibration source C (Cal C)}. This source  will
illuminate all the detector sensitive area to map the gain at one
energy. This source is composed of an ${}^{55}$Fe iron radioactive
source glued in a holder which is screwed in a body. A collimator
allows X-ray photons to impinge on the detector sensitive area
only when the source is in front of the GPD. \item
\textbf{Calibration source D (Cal D)}. This source will illuminate
all the detector sensitive area as does the Cal C, to map the gain
at another energy. Cal D is based on an ${}^{55}$Fe source, glued
in an aluminum holder which illuminates a Si target mounted on a
body to extract K$\alphaup$ fluorescence from silicon. This source
produces fluorescence emission at 1.7 keV. Silicon would provide,
together with Cal C, more \emph{leverage} for gain calibration
and, since photoelectron tracks are smaller at this energy, the
gain map can have a higher spatial resolution. It is worth noting
that the design is such that X-ray photons from ${}^{55}$Fe cannot
directly impinge on the GPD sensitive area to avoid detector
saturation.
\end{enumerate}

All calibration sources inside the FCW contain an ${}^{55}$Fe
nuclide, whose activity naturally decays with a half life of 2.7
years. In Table \ref{tab:FlightSource} we report the expected
flight rate at the beginning of the operative phase of IXPE that,
at the time of writing, is on November 2021.

\begin{deluxetable*}{ccc} [ht]
\tablecaption{The nominal activity at the beginning of life and
the averaged counting-rate expected at the beginning of the IXPE
operative life, assumed to be on November 2021}
\label{tab:FlightSource} \tablewidth{0pt}
\tablehead{\colhead{Calibration Source} & \colhead{Activity (mCi)}
& \colhead{Counting rate (c/s)}} \startdata \textbf{Cal A} & 100 &
1.4 @ 3.0 keV \\
& & 17 @ 5.9 keV
\\ \textbf{Cal B} & 20 & 48 @ 5.9 keV \\
\textbf{Cal C} & 0.5 & 90 @ 5.9 keV \\ \textbf{Cal D} & 100 & 76 @
1.7 keV
\enddata
\end{deluxetable*}

Replacement/installation of the radioactive nuclide is possible
through a dedicated opening in the DU lid. Radioactive sources,
glued in their holders, can be extracted with dedicated tools and
replaced with new ones in new holders.

\section{Detector Service Unit}
\label{sec:DetectorServiceUnit} Thanks to the functionalities of
the ASIC and of the GPD readout system, most of the necessary
operations on the data are performed by the DAQ electronics
\citep{Barbanera2021} including acquisition, digitation of the
pixel charge, zero-suppression and efficient data packing. We
designed the DSU in order to, mainly, perform data formatting in a
way suitable for the storing into the spacecraft memory (5 GByte)
and for using the S-band for downloading. The DSU also controls
the Filter $\&$ Calibration Wheels and manages the payload
operative modes.

The DSU provides the following functions for the operation and
autonomy of the IXPE instrument:
\begin{enumerate}
\item  Power supply generation (secondary power to supply the DU's
sub-units) \item  Telemetry management (data acquisition,
formatting and transmission to the S/C) \item  Commands management
(reception, verification, generation, scheduling, distribution or
execution) \item  Time management (synchronization of the DUs,
Pulse Per Second (PPS) distribution and time tagging) \item Filter
and Calibration wheel management \item Science data management
(retrieving, isolated pixels removal, formatting and transmission
to the S/C) \item Payload mode control \item  GPD temperature
control \item Fault Detection Isolation and Recovery (FDIR)
management
\end{enumerate}

The DSU consists of two boards, with cold redundancy, and one
backplane for internal DSU signal routing. The boards are the
Single Board Computer (SBC) and the Power \& Service Board (PSB).

\begin{figure}[htpb]
\caption{ DSU high level architecture.}
\label{fig:DSUArch}
\centering
\includegraphics[width=7.07in, height=5.83in,keepaspectratio=true]{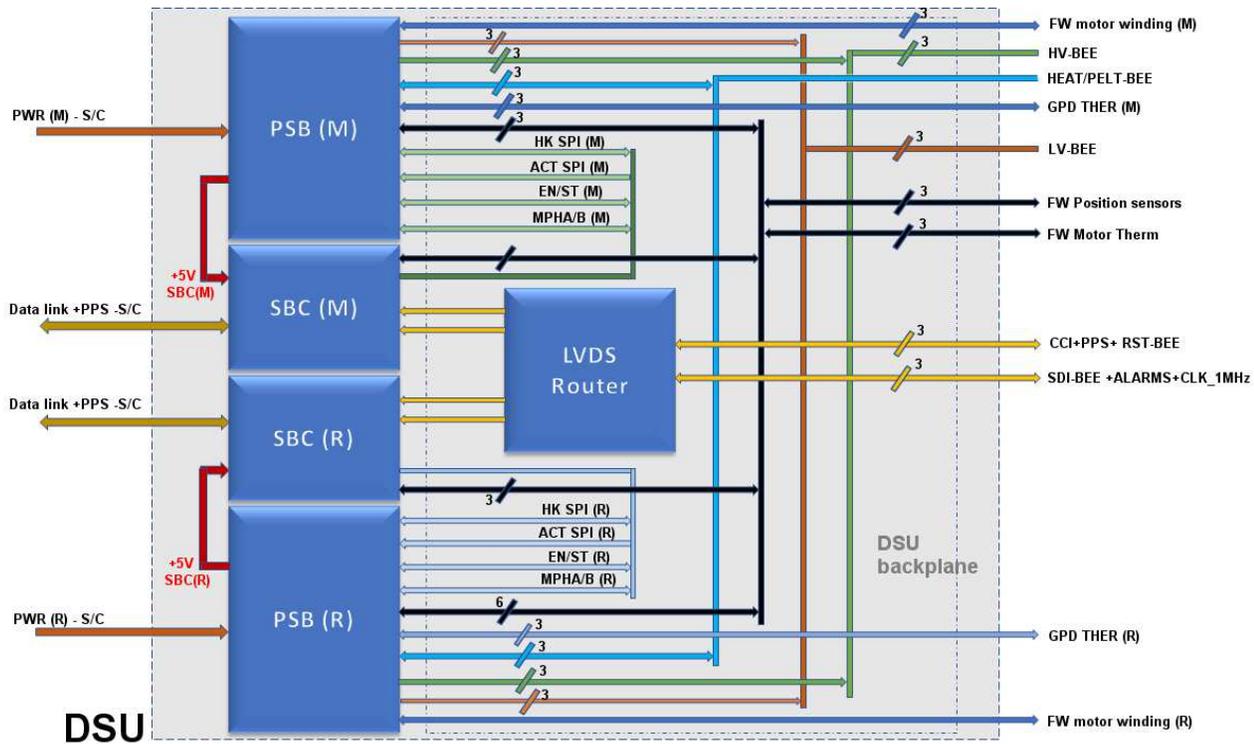}
\end{figure}

Figure \ref{fig:DSUArch} shows the high level architecture of
the DSU with its main electrical interfaces.

The SBC implements the instrument control and configuration and
the instrument data processing and formatting. In fact, it
performs a rejection of raw data called 'orphan removal' (see
Section \ref{sec:OrphRemov}). The SBC is based on a LEON3FT
Central Processing Unit (CPU). The LEON3FT CPU is implemented in
ASIC provided by COBHAM (UT699). A companion Field-Programmable
Gate Array (FPGA), belonging to RTAX family by MICROSEMI, is
included to support the processor operation and to provide the
management of data interfaces required by the IXPE instrument and
not included in the ASIC.

The PSB performs different tasks relevant to power management; in
particular, this board hosts the power converters aimed at the
generation of required supply voltages both for DU and DSU
electronics. The drivers for DU thermal actuators (heaters and
Peltier cells) and for the FCW are also installed. The DSU, also,
implements the signal conditioning for the Resistance Temperature
Detectors (RTD) of the DU temperature control and for the motor
positioning sensors of the FCWs and their temperature.

DSU architecture supports a redundancy philosophy based on
cold-sparing approach. The only exception is the power I/F for
each FCW motor that is equipped with redundant coil for each motor
phase. For each motor phase, the main coil is connected to the
main PSB motor driver and the redundant coil to the redundant PSB
motor driver. For this reason, one temperature sensor is connected
to the main PSB and the other one to the redundant PSB.

The data interface between DSU and each DU consists of the
following 5 sets of three signal lines which comply with
Low-Voltage Differential Signal LVDS standard:

\begin{enumerate}
\item  Three Command and Communication Interfaces (CCI) for
command transmission from DSU to each DU

\item  Three Serial Data Interfaces (SDI) for sensor data
transmission from each DU to DSU

\item  Three reset lines, one for each DU

\item  Three input alarm lines, one for each DU, used to switch
off the relevant DU in case of anomaly detected by the BEE

\item  Three output timing signals for DU internal acquisition
timing verification
\end{enumerate}

IXPE uses a centralized spacecraft memory bank for data storage
and transmission. The memories on-board the DSU are used only to
store the boot SW and support the CPU and FPGA operation.

The local bus connects the LEON3FT CPU, the FPGA, the central SRAM
(Static Random Access Memory), the PROM (Programmable Read Only
Memory) and MRAM (Magnetoresistive random-access memory). In order
to minimize bus loading, the MRAM is connected to the local bus
via bus switches. After boot phase completion, the MRAM is
disconnected from the bus.

The DSU interfaces the Spacecraft with two (one Nominal and one
Redundant) unregulated power lines on separate connectors. The
nominal voltage of the power lines is 28 V. Minimum and maximum
values of input power provided by S/C is respectively 26 V and 34
V.

\subsection{IXPE instrument timing architecture}
\label{sec: InstrumentTimingArchitecture} 
\begin{figure}[ht]
\caption{The Architecture of timing for the instrument.}
\label{fig:Timing}
\centering
\includegraphics[width=8.0in, height=4.00in,keepaspectratio=true]{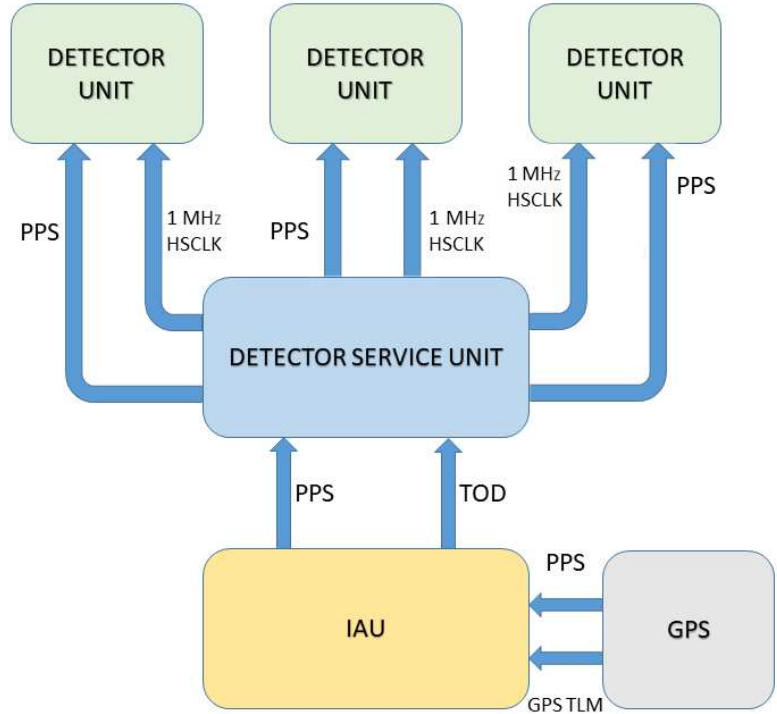}
\end{figure}

The IXPE satellite
manages the timing of the events using a GPS receiver. The GPS
provides a PPS received by the instrument through a dedicated
line. The Spacecraft also provides the Time of the Day (TOD) with
a frequency of 1 Hz. The TOD contains information about the
validity of the previously provided PPS and the time of the next
PPS. A plot showing the timing scheme is shown in Figure
\ref{fig:Timing}.

The DSU is equipped with a temperature compensated crystal
oscillator (TCXO) with an accuracy $\mathrm{<}$ 4ppm. The PPS and
the 1 MHz clock generated by the 1 MHz local oscillator are
provided to each DUs.

The On-Board Time (OBT) management is based on a Master OBT
implemented in the DSU and 3 Local OBTs implemented in the DUs.
The Master OBT and the Locals OBTs are composed of (i) a 29-bit
counter for the seconds ($\mathrm{>}$ 16 years of mission
lifetime) and (ii) a 20-bit counter for the microseconds (1 $\mu$s
resolution) and a register for the OBT error (error counter).

At boot, the Master OBT is initialized by subtracting the Start
Mission date (stored in the DSU MRAM) from the GPS time.

During nominal observation, when the GPS is valid, the second
counter is incremented upon receiving the PPS and the microsecond
counter is incremented by the 1 MHz local oscillator. At the
arrival of the PPS, the difference of the microsecond counter with
respect to 10$^{6}$ is stored in the Master OBT error counter and
included in the Housekeeping packet data with 1 $\mu$s resolution
and 1 s rate. The Local OBTs are updated in the same way by the
FPGA located in the BEE using the PPS provided by the S/C through
the DSU and the 1MHz clock provided by the DSU.

If the previous PPS is not valid, as indicated by the TOD, the PPS
provided by the S/C is not used. The DSU synthesizes the PPS for
the DUs using the Local TCXO. During this operation mode the OBT
Error is set equal to zero. The non-valid condition information is
included in the Housekeeping (HK) data. The Master OBT remains in
free running until the reception of the sequence of TOD reporting
that the PPS is newly valid. In this case the DSU re-initializes
the Instrument timing as at Instrument boot.

The alignment of the Local OBT with the Master OBT is verified
using the housekeeping telemetry (TM). The HKs Telemetry (TM)
comprises the values of the Locals OBT and the Master OBT every
second. At the MOC/SOC these values are monitored and in case of
misalignment an action recovery is planned (e.g., Time Reset by
telecommand).

\subsection{Orphan removal of data}
\label{sec:OrphRemov}

The fulfillment of the requirement on the maximum counting rate is
possible thanks to the DAQ that packs consecutive zeroes by
writing their number in a single word \citep{Barbanera2021} and
the isolated pixel removal in the DSU. As a matter of fact the
only data processing made on-board by the DSU is the removal of
isolated pixels (called \emph{orphans} in the photoelectron track
image) generated by noise. This removal is performed with a simple
algorithm on the serial stream of the zero-suppressed data
delivered by the DU for each event. It requires a limited memory
allocation and is performed by the FPGA without impact on the CPU
timing. By means of the packing in the DU and the removal of
isolated pixel it is possible to reduce the data rate within the
requirements for the S-band. In fact, only the pixels that
identify the tracks are down-loaded, suppressing the isolated
pixel generated by noise.

\subsection{Instrument Operative Mode Specifications $\&$
Transitions} 
\begin{figure}[ht]
\caption{The operative modes of IXPE payload and the transition
via telecommand and FDIR.} \label{fig:PLModeandTrans} \centering
\includegraphics[width=8.0in, height=4.00in,keepaspectratio=true]{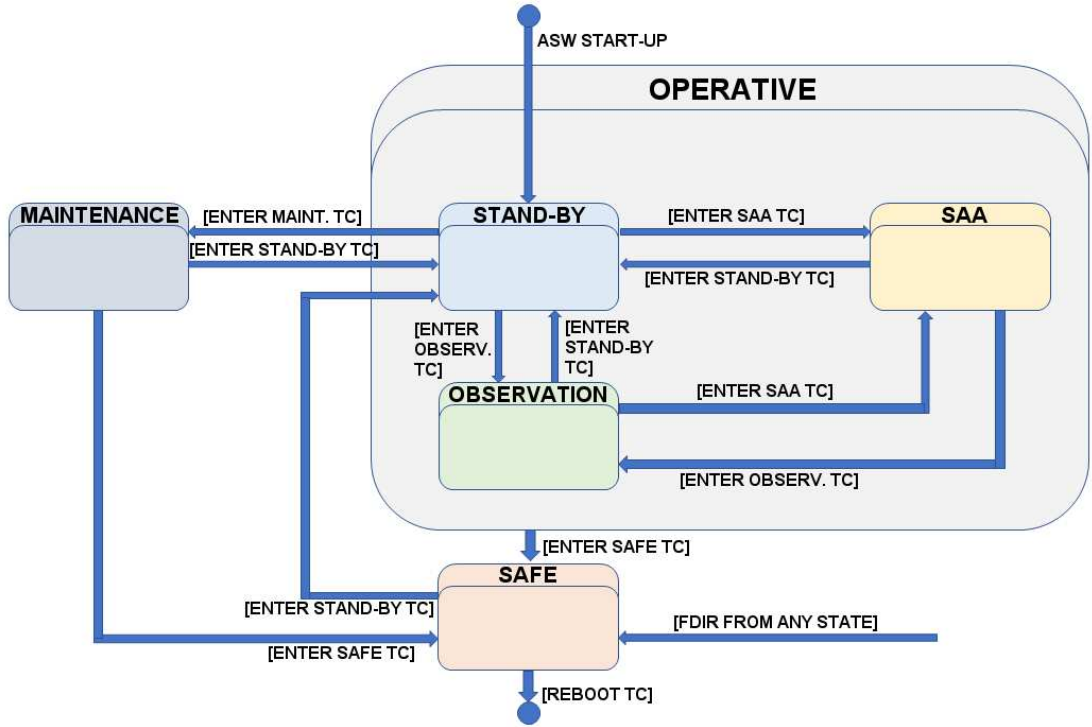}
\end{figure}

The Instrument operative modes are managed by the DSU
and they are selected either via telecommand, or by activating
autonomously a FDIR procedure. The BEE is controlled by the DSU,
and only if it generates an alarm condition, the DSU responds
setting the instrument in Safe mode. The transition between modes
is described by Figure \ref{fig:PLModeandTrans}.

The foreseen payload operative modes are the following:

\begin{itemize}
    \item \textbf{Boot} (transitional mode).
    BOOT is the start-up mode at power on. A limited SW application,
    called Boot Software, runs from the Programmable Read Only Memory (PROM)
    in order to perform all the checks and the initialization of
     instrument resources and items.
    In the nominal case the Application Software (ASW) completes the
    initialization sequence execution by entering the instrument in
    STAND-BY. The boot phase lasts around 60 seconds.

    \item \textbf{Maintenance}.
    This mode is reserved to support the in-orbit maintenance
    program. This mode can be reached on request when the DSU SW mode
    is in Stand-By and is dedicated to memory management operations
    like code and data load/download in/from MRAM and SRAM.
    \item \textbf{Stand-by}.
    Nominally, at the end of the boot-strap phase, the DSU moves in
    Stand-by mode to start the instrument monitoring and control.
    The stand-by mode supports the starting and handling of the thermal
    regulation, powering on and off the instrument items, the
    processing of the incoming telecommands and the generation
    of the related TM.
    Finally it configures the detector units and the science data
    processing.
    \item \textbf{Observation}.
    From the point of view of the DSU, all instrument
    calibration and scientific modes are managed into a unique
    Observation mode that is preventively configured while in
    Stand-by mode. This mode supports the handling of thermal
    regulation, the handling of the Time of Day message, the
    collection of the housekeeping and the generation of scientific
    data.
    \item \textbf{South Atlantic Anomaly} (SAA).
    This mode is used when the satellite crosses the South Atlantic Anomaly
    that happens once per orbit. In this mode, the High Voltage (HV) are ramped-down
    below the voltage that sets multiplication, the science data generation
    is disabled while housekeeping data are generated and collected as usual.
    \item \textbf{Safe}.
    This mode is used before switching-off the instrument or performing a SW reboot
    and managing the FDIR conditions. This mode preserves DUs and HVs status (i.e., no ramp-off or
    switch off are performed on hardware) when entered by telecommand. If entered by
    FDIR, HVs ramp-off, DUs switch off and FCW rotation to CLOSE
    can be performed as part of the \emph{Recovery} action. In this operative mode,
    the instrument generates and collects the housekeepings.
    Only during Safe mode it is possible to perform a reboot of the ASW.
    \item \textbf{Reboot} (transitional mode). At the DSU switch-on, after the BOOT SW activities
     the ASW is loaded into SRAM and executed. The ASW checks the position
     of the Filters and Calibration Wheels and the status of the Detector Units
     and High-Voltages Boards.

     If the FCWs are closed and the HVBs are off, the ASW moves to
     STAND-BY mode. If the FCWs are not closed or if the HVBs are ON,
     the ASW moves the FCWs to CLOSED position, performs the ramp-off procedure if HVBs
     voltages are greater than zero, it switches off the DUs, it moves SW
     operative mode to Safe.
\end{itemize}

\section{The Calibration of the IXPE Detector Units}
\label{sec:ICE} Being a discovery mission, celestial sources are
not available for performing in-flight calibration for
polarimetry. In fact the measurement performed 45 years ago on the
Crab nebula \citep{Weisskopf1978} with a non-imaging detector,
cannot be used as a 'flight' calibrator. Indeed this time interval
is already about 5 $\%$ of its lifetime and the Crab-nebula was
found to vary. For these reasons an on-ground detailed calibration
is mandatory.

Furthermore we designed and built an on-board calibration system
aimed at checking variations in the modulation factor and, even if
not with the same accuracy reached on-ground, any variation on
spurious modulation (see Section \ref{sec:FCS})

The calibration of the Detector Units is a key activity for IXPE,
Details of the calibration facility and of the calibration results
will be presented in forthcoming papers. Beside validating the
Monte Carlo prediction for the modulation factor, more importantly
it provides physical parameters like (i) the low-energy spurious
modulation which is the instrument response to unpolarized
radiation (see Section \ref{sec:GPD} and Figure
\ref{fig:SpurModFact}) and (ii) the gain non-uniformities that
cannot be modelled in advance: they have been calibrated on ground
and included in the data processing pipeline. The calibration
strategy mimics the observation strategy during flight that
foresees dithering of point-sources. In this way, when we
calibrated the entire detector, we obtained the so called Flat
Field (FF) measurements. When we calibrated at high count rate
density only a selected central part of the detector active
region, we obtained the so called Deep Flat Field (DFF)
measurements. The latter are aimed at reaching very large
statistics within the dithering region foreseen in flight for
point-like celestial sources. When we calibrated with a narrow
beam (less than 1-mm diameter), we obtained the so called 'spot
measurements'.

The calibration activity (40 days for each DU) was directed toward
measuring the following physical quantities:
\begin{enumerate}
    \item Spurious modulation (60$\%$ of the calibration time) at 6 energies
    between 2 keV and 5.9 keV in a DFF configuration (central region with 3.25 mm
    diameter at a density of 10$^6$ counts/mm$^2$) and FF configuration
    (all detector sensitive area with 10$^5$c/mm$^2$)
    \item Modulation factor (17.5$\%$ of the calibration time at 7 energies),
    and different polarization angles,
    between 2.0 keV and 6.4 keV in DFF configuration (3.25 mm
    radius, 10$^4$~counts/mm$^2$) and in FF configuration (7 mm
    radius with the same statistics of DFF but in a larger area)
    \item Efficiency at, at least, three energies (2.5$\%$ of the calibration
    time)
    \item Spatial resolution (2.5$\%$ of the calibration time) of unpolarized radiation at 3 energies
    in an array of 3 $\times$ 3 spots
    \item Spatial resolution (2.5$\%$ of the calibration time) of polarized radiation at 3 energies in one point
    \item Grey filter transparency at 5 energies, inclined penetration (focusing effect) and calibration of the FCW sources (rest of the fractional time)
\end{enumerate}

 From the above measurements it was possible to
derive, also, the following physical quantities:

\begin{itemize}
  \item Accuracy of the polarization angle measurement at 7
  energies
  \item Map of the gain variations in the active region
  \item Measurement of the energy resolution at different energies
  \item Measurement of the dead time at different energies
\end{itemize}

We measured the modulation factor at different angles not only to
measure the accuracy of the polarization angle but also to check
that spurious modulation is properly subtracted. Indeed any change
of the modulation factor with polarization position angle is
expected in the presence of spurious effects which mimic the
presence of a secondary, albeit small, polarization component.

We organized the calibration in such a way that the time dedicated
to measure spurious modulation at different energies accounts for
60$\%$ of the total calibration time. As a matter of fact, in
order to measure spurious modulation at a level of 0.1$\%$ across
the sensitive area, we collected tens of millions of counts for
each energy. The calibration time was set considering the maximum
counting rate allowed for the DUs, due to their dead time, and the
power of the X-ray tubes.

In order to perform such calibrations we built two stations. One
station is the main calibration station (called ICE, Instrument
Calibration Equipment). The other, called AIV/T Calibration
Equipment (ACE) is designed to illuminate contemporaneously up to
three DUs to support the integration and test of the whole
instrument, but it can host the same calibration sources as the
main station. During some periods, ACE was indeed used to carry
out calibration measurements in parallel with ICE. They are the so
called AIV$\&$T Calibration equipment (ACE1 and ACE2).

The ICE is derived from an earlier design which was used for 15
years at INAF-IAPS \citep{Muleri2008b, Muleri2021}.  In
particular, it includes:

\begin{enumerate}
\item  the X-ray sources used for calibration and functional
tests.  Each source emits X-ray photons at known energy and with
known polarization degree and angle. Polarized sources exploit
Bragg diffraction at nearly 45$^\circ$ from a set of crystals (see
Table \ref{tab:CrystalTubes}) with suitable X-ray tubes.
Unpolarized sources are based on the direct emission from X-ray
tubes or on the extraction of fluorescence emission from a target.
The direction of the beam, the position angle for polarized
sources, and its position can be measured with respect to the GPD
thanks to its fiducial points and the use of a Romer measurement
Arm.

\item  the test detectors, one commercial Silicon Drift Detector
(SDD) by AMPTEK and one X-ray CCD camera by Andor. They are used
to characterize the beam before DU calibration and as a reference
for specific measurements (e.g., the measurement of quantum
efficiency with the SDD).

\item  all the electrical and mechanical equipment required  to
support the DU and the calibration sources, monitor the relevant
diagnostic parameters and assure safety during calibrations.
\end{enumerate}

\begin{figure}[ht]
\caption{The instrument calibration equipment at \emph{INAF-IAPS}.
On the left a flight DU under calibration with a polarized source.
At the center the Romer arm used for metrology is visible. On the
right the CCD to study the beam image and the SDD to study the
beam spectrum are just visible.} \label{fig:ICE} \centering
\includegraphics[width=4.0in, height=3in,keepaspectratio=true]{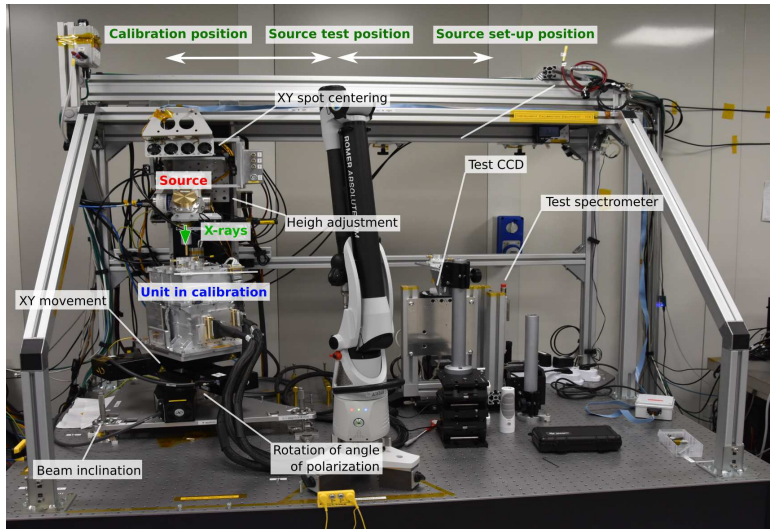}
\end{figure}

The ICE in the configuration with the polarized source is shown in
Figure \ref{fig:ICE}. The DU is mounted in the ICE without the
stray-light collimator and the UV filter. In this way we minimize
the distance between the X-ray source and the GPD and hence air
absorption and beam divergence. The DU is placed on the top of a
tower that allows to:
\begin{enumerate}
\item  move the DU on the plane orthogonal to the incident beam
with an accuracy of $\mathrm{\pm}$2 $\muup$m (over a range of 100
mm) to map the GPD sensitive surface. \item  rotate the DU on the
plane orthogonal to the incident beam with an accuracy of
$\mathrm{\pm}$7 arcsec, to test the response at different
polarization angles and to average residual polarization of
unpolarized sources, if necessary. \item  tip/tilt align the
orthogonal direction of the GPD to  the incident beam. Two out of
the three feet of the tip/tilt plate are manual micrometers, but
one is motorized to carry out automatically measurements with the
beam off-axis of a series of known angles, between $\mathrm{<}$1
degree and about 5 degrees, e.g., to simulate the focusing of
X-ray mirror shells.
\end{enumerate}

\begin{deluxetable*}{ccccccc}
\tablecaption{The Crystals and the X-ray tubes used during GPD
acceptance and DUs calibration. This is a subset of available
energies. The column $\frac{R_{p}}{R_{s}}$ represents the ratio of
the integrated reflectivity for polarization parallel and
perpendicular to the crystal lattice plane at that given energy.
From this ratio it is possible to derive the polarization degree
of the reflected beam (from \cite{Henke1993})}
\label{tab:CrystalTubes} \tablewidth{0pt}
 \tablehead{
\colhead{Crystal}&\colhead{X-ray tube} & \colhead{Energy} &
\colhead{2d} &
\colhead{Diffraction} & \colhead{$\frac{R_{p}}{R_{s}}$} & \colhead{Polarization}\\
\colhead{} & \colhead{} & \colhead{[keV]}& \colhead{$\AA$} &
\colhead{Angle}&
\colhead{}&\colhead{[$\%$]}\\
\colhead{}&\colhead{}& \colhead{}&
\colhead{}&\colhead{[deg]}&\colhead{}&\colhead{}} \startdata
PET (002) & Continuum & 2.01 & 8.742 & 45.0 & 0.0 & $\approx$100.0\\
InSb (111) & Mo L$_{\alpha}$ & 2.29 & 7.481& 46.361& 0.0034&99.2$\%$\\
Ge (111) & Rh L$_{\alpha}$ & 2.7 & 6.532 & 44.877 & 0.00 &
$\approx$100.0\\
Si (111) & Ag L$_{\alpha}$ & 2.98 & 6.271 & 41.562 & 0.0252 &
95.1$\%$ \\
Al (111) & Ca $K_{\alpha}$ & 3.69 & 4.678 & 45.909 & 0.0031 &
99.4$\%$\\
Si (220) & Ti K$_{\alpha}$ & 4.51 & 3.840 & 45.716 & 0.0023 &
99.5$\%$\\
Si (400) & Fe K$_{\alpha}$ & 6.40 & 2.716 & 45.511 &  & $\sim$ 100 $\%$\\
\enddata
\end{deluxetable*}

The calibration of the flight DUs was performed successfully
following the on-ground calibration plan. The major effort was
 the calibration of the modulation factor and the
characterization and the subtraction of spurious modulation at
different energies. The detailed description of these features
is the subject of forthcoming papers.

\section{Instrument performance}
\begin{figure}
\caption{\textbf{a}. Spurious modulation at instrument level (on a
spot of 1.5 mm radius and combining the three DU's clocked at
120$^\circ$). \textbf{b}. Modulation factor of the instrument
after spurious modulation subtraction (Average of the three DUs),
with dithering on a flight-representative region of 1.5 mm
radius.} \label{fig:SpurModFact}
          \gridline{\fig{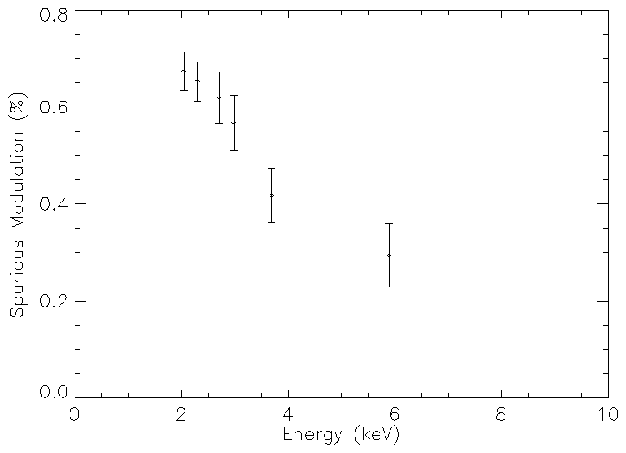}{0.5\textwidth}{(a)}
          \fig{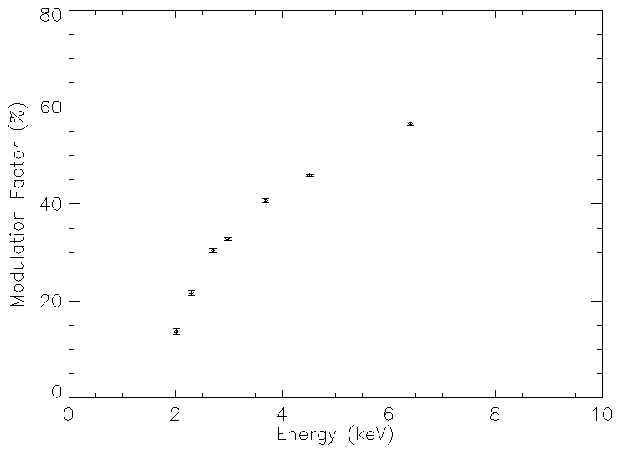}{0.5\textwidth}{(b)}
          }
\end{figure}

We summarize here the instrument performances after calibration.
We show in Figure \ref{fig:SpurModFact} the results for spurious
modulation and modulation factor. In the left panel we show
spurious modulation re-phasing the modulation curves when
considering a dispositions of the DUs as in flight (120$^\circ$
clocking around Z-axis). The orientation of the DUs allows for
reducing the spurious modulation of the instrument with respect to
that of a single DU. Such spurious modulation is subtracted on
event by event base. The right plot shows, instead, the modulation
factor after keeping 80\% of the data among the most elongated
track and event-by-event spurious modulation subtraction.


In Table \ref{tab:InstrumCharact} we show the characteristics of
the Instrument as average of the three DUs at two characteristic
energies. These parameters can be considered as representative of
the entire instrument. The detailed calibration results for each
flight DU will be presented in a set of forthcoming papers.

\begin{deluxetable*}{ccc}
\label{tab:InstrumCharact} \tablecaption{Instrument performance.
Spurious modulation is effectively measured at instrument level,
thus considering 120$^\circ$ clocking of three DUs. Data are
extracted in a circular region of 1.5 mm radius. Efficiency
includes only the GPD contribution extrapolated at the time of the
flight. \label{tab:InstPerf}} \tablewidth{0pt} \tablehead{
\colhead{Parameter} & \colhead{2.69 keV} & \colhead{6.40 keV}}

\startdata Modulation factor & 30.4 $\%$  $\pm$ 0.4 $\%$ & 56.6 $\%$ $\pm$ 0.4 $\%$ \\
 Efficiency (expected during flight) & 13.5$\%$ & 1.7$\%$\\
 Gray filter transparency & 17.4 $\%$ & 88.0 $\%$ \\
 UV Filter transparency & (95.94 $\pm$ 0.26)$\%$ & (99.44 $\pm$ 0.38) $\%$ \\
 Spurious modulation & (0.62 $\pm$0.05) $\%$  & (0.29 $\pm 0.06$) $\%$ (@5.89 keV)\\
 Systematic error on the P.A. determination & (0.143$$ $\pm$ 0.094)$^\circ$ & (0.186 $\pm$ 0.097)$^\circ$\\
 Energy resolution & (22.2$\pm$0.5)$\%$ & (16.3$\pm$0.1)$\%$  \\
 Position resolution (HPD) & (118.7 $\pm$ 5.2) $\mu$m & 120.0 $\pm$ 5.8) $\mu$m\\
 Dead Time & 1.1 ms &  1.2 ms \\
 \hline
 \colhead {Parameter} & Value \\
 \hline
 Timing accuracy & 1-2 $\mu$s (by means of the use of the PPS)\\
 Timing resolution & 1 $\mu$s \\
 Common FoV  &  9' \\
 Dithering radius & 1'.6 standard (or 0'.8 or 2'.6 or no-dithering)\\
 Expected background rate (2-8 keV) & 1.9 10$^{-3}$ c/s/cm$^{2}$/keV\\
 Expected Crab (nebula + pulsar)\footnote{Crab spectrum as in \cite{Zombeck2007}} rate (2-8 keV) & 150 c/s \\
 Expected Crab (nebula + pulsar) rate (1-12 keV) & 240 c/s \\
\enddata

\end{deluxetable*}

\section{Processing Pipeline}
\label{DataFormat} Data are analyzed on ground. The first step of
the pipeline is to convert binary telemetry files to FITS standard
format by means of a parser. At this stage, data contain all the
information provided by the Instrument and the relevant
information from the spacecraft. In particular they contain the
raw, pedestal-corrected, track images, the time of each count and
all the relevant house-keeping like temperatures and
event-by-event dead time. The FITS-converted data are then
processed in order to transform the engineering unit in physical
units, removing the non-necessary engineering information and
including Good Time Interval information. The processing pipeline,
then, uses calibration data in order to (1) equalize the pixel
response of the track image; (2) identify by means of a clustering
algorithm the photoelectron (principal) track; (3) estimate the
emission direction in detector coordinates; (4) estimate the
location of the events in detector coordinates; (5) estimate the
energy of the track by applying (a) correction for charging, (b)
correction for temperature and (c) GEM gain non-uniformities. The
last step of the correction is the removal of spurious modulation,
which is applied on the Stokes parameter calculated for each
event. At this stage, track images are removed and data contain
the position in detector coordinates, the time of the event, the
energy of the event in ``pulse invariant'' unit, the emission
direction in detector coordinates and the Stokes parameter of the
event according to \cite{Kislat2015} definition with only an extra
factor of 2. Next step is to merge the detector data from the
three DUs, including their clocking at 120$^\circ$, and to convert
to celestial coordinates. The final event file to be used for
science analysis, after that all the procedures are applied,
contains all the information typical of imaging X-ray Astronomy
mission, but with the addition of the information on the emission
direction of the photoelectron in celestial coordinates.

\section{Conclusion}
\label{Conclusions} We presented the instrument that was
successfully designed, built, calibrated and installed in the
spacecraft for the IXPE mission. These activities spanned 3.5
years, including phase B and the development time is compliant
with readiness for a launch starting 17 November 2021. We reviewed
the instrument design and the results of the calibration and we
compared them with the ones expected on the basis of scientific
astrophysical requirements. The modulation factor, the energy
resolution, the position resolution of the instrument all meet the
scientific requirements. The quantum efficiency is below
requirement, but when factoring in all other parameters, the top
level polarization sensitivity requirement is still met.
Background also is expected to be slightly higher than the
requirement \citep{Xie2021}. However, due to the relatively bright
sources in the IXPE observing plan, we anticipate no effect in
point-like sources observation and only a mild effect is expected
in low surface brightness extended sources like the molecular
clouds in the vicinity of the Galactic Center or in SN1006
supernova remnant. The calibration of the spare telescope, the
only one to be calibrated, has been accomplished and its results
will be part of a forthcoming paper. The preliminary results are
such that in-flight performance of the IXPE payload will permit a
new window to be opened in x-ray polarimetry.

\acknowledgments The Italian contribution to the IXPE mission is
supported by the Italian Space Agency (ASI) through the contract
ASI-OHBI-2017-12-I.0, the agreements ASI-INAF-2017-12-H0 and
ASI-INFN-2017.13-H0, and its Space Science Data Center (SSDC), and
by the Istituto Nazionale di Astrofisica (INAF) and the Istituto
Nazionale di Fisica Nucleare (INFN) in Italy. The italian IXPE
collaboration acknowledges the support of Martin C. Weisskopf,
Brian D. Ramsey, Stephen L. O'Dell, Allyn Tennant, Wayne H.
Baumgartner, Jeff Kolodziejczak and Stephen D. Bongiorno of
NASA-MSFC in the development of the IXPE instrument by means of
continuous and useful discussions and for carefully reading this
manuscript.

\bibliographystyle{aasjournal}
\bibliography{../../../../../../ReferenceBib/References}{}



\end{document}